\documentclass[12pt]{article}

\usepackage{scicite}
\usepackage{graphicx}
\usepackage{times}
\usepackage{comment}

\usepackage[font=small,labelfont=bf]{caption}

\makeatletter
\newcommand*{\addFileDependency}[1]{
  \typeout{(#1)}
  \@addtofilelist{#1}
  \IfFileExists{#1}{}{\typeout{No file #1.}}
}
\makeatother

\usepackage{xr} 
\usepackage{lineno}

\usepackage{amsmath,amssymb}

\usepackage{booktabs}

\usepackage[capitalize]{cleveref} 

\usepackage{url}

\newenvironment{sciabstract}{%
\begin{quote} \bf}
{\end{quote}}

\title{Human mobility in the metaverse} 

\author
{Kishore Vasan$^{1}$, Márton Karsai$^{2,3}$, Albert-László Barabási$^{1,2,4\ast}$\\
\\
\normalsize{$^{1}$Network Science Institute, Northeastern University}\\
\normalsize{$^{2}$Department of Data and Network Science, Central European University}\\
\normalsize{$^{3}$National Laboratory of Health Security, Alfréd Rényi Institute}\\
\normalsize{$^{4}$Department of Medicine, Brigham and
Women’s Hospital, Harvard Medical School}\\
\\
\normalsize{$^\ast$To whom correspondence should be addressed; E-mail:  a.barabasi@northeastern.edu.}
}

\date{}

\begin{document} 

\baselineskip24pt


\maketitle 


\begin{sciabstract}
The metaverse promises a shift in the way humans interact with each other, and with their digital and physical environments. The lack of geographical boundaries and travel costs in the metaverse prompts us to ask if the fundamental laws that govern human mobility in the physical world apply. We collected data on avatar movements, along with their network mobility extracted from NFT purchases. We find that despite the absence of commuting costs, an individual’s inclination to explore new locations diminishes over time, limiting movement to a small fraction of the metaverse. We also find a lack of correlation between land prices and visitation, a deviation from the patterns characterizing the physical world. Finally, we identify the scaling laws that characterize meta-mobility and show that we need to add preferential selection to the existing models to explain quantitative patterns of metaverse mobility. Our ability to predict the characteristics of the emerging meta-mobility network implies that the laws governing human mobility are rooted in fundamental patterns of human dynamics, rather than the nature of space and cost of movement. 

\end{sciabstract}

\section*{Introduction}

With the ambition to create seamless, immersive, and personalized experiences by integrating augmented and virtual reality, artificial intelligence, and blockchain technology\cite{caldarelli2023role}, the metaverse promises a paradigm shift in the way humans interact with each other, as well as with their digital and physical environments, and offers new ways to participate in virtual economies\cite{ball2022metaverse}. Driven by the potential of the technology and its far-reaching impact on society, in 2022, metaverse companies attracted over \$120 billion in investments\cite{mkinsey}. Thanks to this influx of attention and resources, the metaverse already enables individuals to explore three dimensional (3D) virtual worlds\cite{gadekallu2022blockchain}, transfer digital assets through cryptographic safeguards\cite{marthews2019blockchain, weyl2022decentralized}, create and trade Non-Fungible Tokens (NFTs)\cite{vasan2022quantifying,nadini2021mapping}, and unlock novel avenues for self-expression\cite{gomez2023promise}. 

Historically, human movements are confined to the tangible realm of the physical world, limited by geographical and natural boundaries and the time and cost associated with commuting and movement. Facilitated by detailed data on individual human movements, studies have unveiled the existence of multiple universal laws and patterns governing human mobility\cite{gonzalez2008understanding, yuan2012discovering, zhao2021characteristics}, and have inspired the development of a family of quantitative models capable of explaining these patterns\cite{simini2012universal, crandall2010inferring}. The resulting modeling framework quantifies the balance of physical proximity and economic prospects in shaping human mobility\cite{eagle2010network, davis2009avatars}, and its impact on socio-economic segregation\cite{schelling1971dynamic, echenique2007measure, netto2015segregated, moro2021mobility, napoli2023socioeconomic}. These models have taken a leading role during the COVID-19 pandemic, when mobility-based approaches played a key role in predicting the spread of the pathogen, proposed restrictions to curb its spread, and have fueled contact tracing algorithms\cite{bonaccorsi2020economic, kojaku2021effectiveness, chang2021mobility, santana2023covid}. 

A distinguishing feature of the metaverse is that individuals are no longer limited by geographic constraints, nor by the expenses and time investment associated with travel, allowing them to seamlessly transition between different locations\cite{szell2012understanding, hu2018life, hazarie2020uncovering}. In this new multi-dimensional landscape, traditional modeling frameworks inspired by transportation modes and residential choices\cite{zhong2017revealing}, opportunities and geographic constrains deeply rooted in the physical space of human existence\cite{yan2017universal, radicchi2012rationality}, may no longer apply. Therefore, we need to ask whether the well-established laws and patterns governing human mobility continue to apply, and if not, how do we develop a new modeling framework for the virtual space, where choices are driven by individual preferences, popularity, and network effects, rather than physical distance. 

Here, we address these questions by exploring data collected from two separate metaverse systems: (a) The first captures individual movements within a two dimensional (2D) virtual environment. As the movements are confined to a pre-defined grid, it offers an environment whose metrics are comparable to the physical world. (b) The second data focuses on contract mobility, capturing the movement of the individuals observed in (a) across different blockchain-based NFT platforms. This data describes mobility in an infinite dimensional space, best described as network mobility. By examining the movements of the same individuals in these two different settings, we can offer a comprehensive perspective on human behavior within the metaverse ecosystem (Fig. \ref{fig:global_mobility} A).

Our results indicate that despite the absence of commuting costs, an individual’s inclination to explore new locations diminishes over time. As a result, movement is heavily concentrated in a few locations in the metaverse, resulting in a pronounced disparity in visitor distribution across various locations. Importantly, we find a lack of correlation between land prices and visitation, underscoring the distinctive feature of the metaverse, where economic and spatial mobility dynamics deviate from the persistent patterns identified in the physical world. Finally, we show that to explain the observed metaverse mobility patterns and the emerging scaling laws, we need to modify the prevailing mobility models, incorporating a preferential selection mechanism for location selection. The proposed modification, a meta-mobility model called \textit{m-EPR}, predicts mobility network and heterogeneous mobility flows in line with the empirically observed features in the metaverse. The ability of the \textit{m-EPR} model to accurately describe the mobility network implies that individual movement in the metaverse is driven by popularity-based dynamics, a feature absent from human mobility in the physical world.

\textbf{Data Collection and Curation}

\textbf{Virtual world mobility.} To examine mobility in the metaverse we focused on Decentraland, one of the first decentralized virtual worlds launched in 2019 (Fig. \ref{fig:global_mobility} B), featuring 90,601 distinct lands (locations), each of size 16 by 16 meters in the virtual space and organized in a symmetrical layout. As the locations are implemented as NFTs on the Ethereum blockchain, we can identify the ownership of each land (Supplementary Section 1.2). We collected temporal mobility data of users on the platform in two separate sessions: from March 15, 2022 to August 7, 2022 (D1) and from August 8, 2022 to September 19, 2022 (D2), together capturing information on 163,770 users and their 251,643,262 movements. Using this data, we build a  virtual world mobility network, where nodes are locations and a link signifies movement across the two locations. 

\textbf{Network mobility}. Every NFT purchase is publicly documented on the blockchain, establishing a fail-proof system of verifiable ownership. Focusing on the set of 163,770 individuals whose mobility we tracked on Decentraland, we extracted their network mobility by collecting data regarding their NFT purchases from two blockchains: (a) \textit{Ethereum}, finding 1,165,310 NFTs from 23,827 contracts (platforms) collected by 14,732 (9\%) users, and (b) \textit{Polygon}, finding 3,112,300 NFTs from 54,918 contracts (platforms) engaging 41,870 (25\%) users. Using this data, we create a contract mobility network, whose nodes are NFT contracts (locations) and edges indicate the subsequent purchases of NFTs from two contracts (Fig. \ref{fig:global_mobility} C).

The decentralized architecture of the Decentraland metaverse and the transparent nature of blockchain technology provide a robust foundation for data integrity and privacy. As the metaverse is run using blockchain, data regarding individual behavior does not contain any privately identifiable information, making it a viable and suitable source for scientific studies\cite{bainbridge2007scientific}. Our study received the necessary approval from the Institutional Review Board (IRB) (Reference number 23-03-29). 

\section*{Results}

\subsection*{Emergence of collective attention}

The globalization of transportation has increased interest in selected spots around the world. For example, the Central Park in New York City, once a local park, now attracts an influx of visitors from all over the world\cite{bertaud2021order}. To test whether patterns of high visitation emerge in the metaverse, which lacks traditional transportation systems and travel restrictions\cite{guimera2005worldwide}, we examine the total number of users that visit a specific location. We find that 76\% (69,433) of the lands received at least one visitor, and there was substantial variation in visitation numbers across these visited locations: 21\% (15,193) received a single visitor, 14\% (9,391) were visited by two visitors, and the remaining 65\% (44,849) received visits by three or more individuals (Supplementary Fig. \ref{fig:si_subset_exploration}). 

Similarly, on the Ethereum blockchain, 31\% (7,524) contracts received only one visitor, 14\% (3,379) received two visitors, and the remaining 54\% (12,924) received three of more visitors. Overall, in the virtual world, the top 1\% of the lands attracted 94\% of all visitors, while on the Ethereum blockchain, the top 1\% of the contracts attracted 77\% of the users, and on the Polygon blockchain, the top 1\% of the contracts attract 96\% of all users, a rather remarkable concentration of visits, unparalleled in the physical world.

To further quantify these patterns, we measured the distribution of the number of visitors to a land or contract ($S$), finding that it is well-approximated by the power law, $P(S) \propto S^{-\alpha}$, where the visitation exponent $\alpha$ in the virtual world is $\alpha_{D1} = 1.98$, and $\alpha_{D2} = 1.92$ (Fig. \ref{fig:individual_mobility} A), and in the contract mobility space $\alpha_{ethereum} = 2.02$ and $\alpha_{polygon} = 2.07$ (Fig. \ref{fig:individual_mobility} B). We find the distributions to be stable over time (Supplementary Fig. \ref{fig:si_pa_evidence}), indicating that in the metaverse a few lands and contracts consistently attract most of the users while most lands and contracts struggle to find visitors.

The price of a land or property in the physical world is known to be influenced by its geographic location, as well as its ability to attract significant foot traffic\cite{west2018scale}. Somewhat surprisingly, we find that the selling price of a land in metaverse is not correlated with the number of visitors it receives ($\beta = 0.002$, CI:[-0.02, 0.02], Supplementary Fig. \ref{fig:visitor_sale_price_land_si}). For example, the land at location (118,-10) attracted 196 visitors and fetched a selling price of \$4,029, while a nearby land (120,-12) attracted 2,053 visitors but was sold only \$3,566 (Supplementary Fig. \ref{fig:price_visitation_example}). In other words, despite receiving an order of magnitude more visitors compared to (118,-10), land (120,-12) was unable to demand a higher valuation. 

Further, urban centers, representing geographical clusters that attract high number of visitors, illustrate the significance of geographical proximity in land visitation patterns\cite{hasan2013understanding}, and its impact on pricing\cite{zukin2009new}. In the virtual world, we discover the emergence of local neighborhoods characterized by high visitation, mirroring patterns observed in the physical world (Supplementary Fig. \ref{fig:neigh_visitors_si} A). However, in the metaverse land prices are not geographically clustered ($\beta = 0.294$, CI:[0.27, 0.31], $R^{2} = 0.1$, Supplementary Fig. \ref{fig:neigh_sale_price_land}), suggesting that proximity to popular places is not a strong determinant of land prices (Supplementary Section 2). This lack of a correlation between land prices and visitation underscores the distinctive feature of the metaverse, where economic and spatial mobility dynamics deviate from the urban development theories and the persistent patterns identified in the physical world \cite{hong2020universal}. 

In summary, our empirical observations reveal a pronounced disparity in visitor distribution across various locations within the metaverse. The spatial layout of locations in the virtual world leads to the emergence of clusters of high visitation, revealing high similarities to the emergence of hotspots in the real-world. 

\subsection*{Characterizing individual movements}

The movement of individuals in the physical world are guided by established patterns of regularity, shaped by the location of home and workplace, and the physical distances between them\cite{alessandretti2020scales,barbosa2018human,bokanyi2021universal}. These patterns give rise to locations characterized by a higher visitation probability and diminishing rates of visitation to new locations\cite{gonzalez2008understanding}. In the metaverse, the absence of traditional places of interest and commuting costs prompts us to ask: do individuals typically explore a significant portion of the available locations? Importantly, do the principles that govern individual mobility in the physical world hold true within the metaverse? 

To answer this, we first measure the number of unique locations visited by an individual. We find that, a typical individual explored 18 locations (lands) within the virtual world, or less than 1\% of all lands (Supplementary Figure \ref{fig:s_explored_si} A). Individuals with ownership of virtual land or digital currency in the metaverse explored an average of 39 locations, while those lacking any financial involvement explored just 14 locations on average. In a similar fashion, an average user bought NFTs from 20 different contracts on Ethereum, representing less than 0.1\% of all contracts, and an average user on Polygon purchased from 8 contracts, representing less than 0.1\% of all contracts (Supplementary Fig. \ref{fig:s_explored_si} B). These patterns suggest that despite the lack of physical and time restrictions to discover and explore new locations, individuals tend to focus their mobility to a small fraction of the metaverse. 

\begin{figure}[ht]
\centering
\includegraphics[width=\linewidth, height = 28em, keepaspectratio]{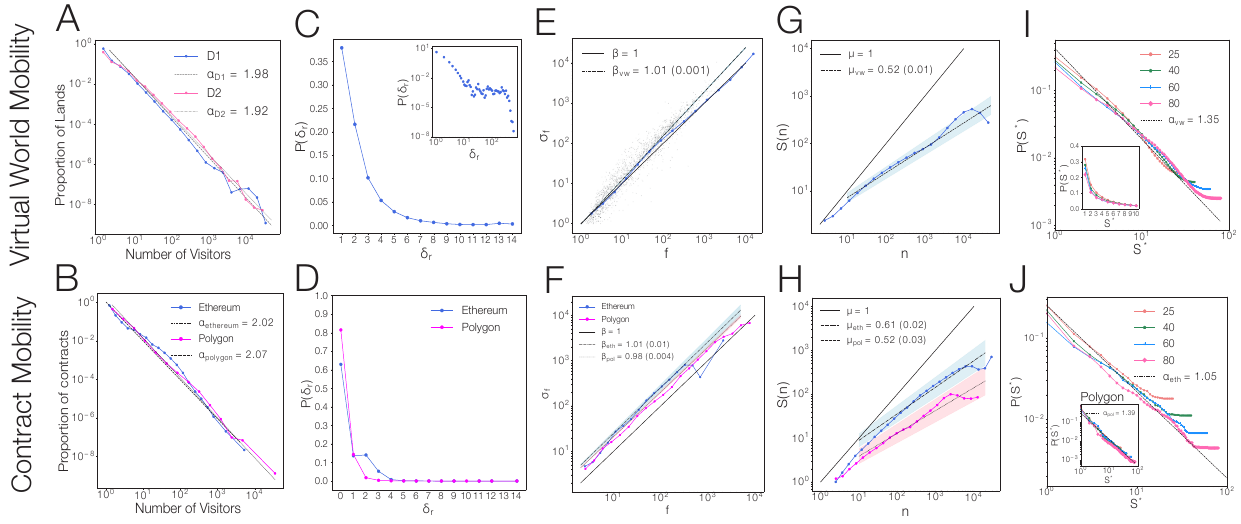}
\caption{\textbf{Individual mobility patterns.} \textbf{(A)} Distribution of number of visitors in the virtual world. We find that the proportion of visitors to different locations (lands) can be well-approximated by $P(S) \propto S^{-\alpha}$, where $\alpha_{D1} = 1.98$ and $\alpha_{D2} = 1.92$. 
\textbf{(B)} Distribution of number of users per node of the network. We find a fat-tailed distribution in number of visitors by contract, well-approximated by the power law exponent $\alpha_{Ethereum} = 2.02$ and $\alpha_{Polygon} = 2.07$. 
\textbf{(C)} Distance travelled in each displacement. We measure the jump distance, $\delta_r$, as the Manhattan distance between two parcels. We find that individuals rarely move past a distance of 10. 
\textbf{(D)} Contract jump distance. We calculate jump distance, $\delta_r$, as the shortest path length between two contracts in the network. A distance of $\delta_r = 0$ indicates purchase of a new NFT from the same contract. We find that the majority of the jumps occur in short distances, irrespective of the blockchain. 
Time allocation at different \textbf{(E)} lands and \textbf{(F)} contracts. We compare the mean visitation frequency $f_{i} = n/S$, where $n$ is the total time spent and $S$ is the number of locations, to the dispersion in visitation counts across all locations, $\sigma_f$. We find that $\sigma_f \propto f^{\beta}$, following $\beta_{vw}= 1.01$ in the virtual world, and $\beta_{ethereum} = 1.01$; $\beta_{polygon} = 0.98$ in the network space. As $\beta \sim 1$ it suggests that as individuals explore more, they tend to unevenly distribute their time across all locations.  
Number of unique locations visited over time. We measure the number of unique locations visited ($S(n)$) as a function of number of steps taken ($n$). We find that $S(n)\propto n^{\mu}$ scales as \textbf{(G)} $\mu_{vw} = 0.52$ in the virtual world and \textbf{(H)} $\mu_{ethereum} = 0.61$, $\mu_{polygon} = 0.52$ in the network space. These insights reveal a sub-linear scaling in the exploration new locations, suggesting that an individuals' inclination to visit more land decreases over time. \textbf{(I)} Location preference and time spent. We rank the locations visited based on the total number of visits to those locations, and display the proportion of time spent at each ranked location in the virtual world. This relationship is well-approximated using a power law with exponent $\alpha_{vw} = 1.35$. Inset shows the same plot in the linear scale. 
\textbf{(J)} Proportion of NFTs purchased from different contracts. Main panel shows the ranked frequency of locations based on the Ethereum blockchain, and the inset shows the results based on the Polygon blockchain. In both systems, the distribution of preference is well-approximated using a power law ($\alpha_{ethereum} = 1.05$, $\alpha_{polygon} = 1.39$).}
\label{fig:individual_mobility}
\end{figure}

To study the role of displacements in the metaverse, we next quantify the jump distance, $\delta_r$. In the virtual world, we can rely on the Manhattan distance, where a distance of five indicates displacement to a land located five units away in any direction (Fig. \ref{fig:global_mobility} B, Supplementary Fig. \ref{fig:si_xy_mobility}, Supplementary Section 3). We find that individuals continue to prefer to move in smaller distances and rarely display large displacements (Fig. \ref{fig:individual_mobility} C). Specifically, jumps greater than a distance of ten accounts for only 18\% of all displacements (Supplementary Fig. \ref{fig:pr_teleport_days}). In contrast to the physical world where distance dictates the visitation frequency between locations\cite{zhong2017revealing, barbosa2018human, schlapfer2021universal}, in the virtual world spatial separation between locations does not affect an individuals likelihood to visit a location (Supplementary Fig. \ref{fig:delta_r_delta_k}). 

In the contract mobility space, distance is measured using the shortest path length between the two contracts (locations) on the contract network (Fig. \ref{fig:global_mobility} C). Similar to the mobility patterns in the virtual world, we find that individuals prefer to purchase from contracts within one to three steps in the contract network (Fig. \ref{fig:individual_mobility} D). Furthermore, individuals tend to repeatedly return to the same contract: we observe a 63\% retention rate on Ethereum, and a 82\% retention rate on Polygon, suggesting a "lock-in" effect in the metaverse, similar to the patterns found in web browsing\cite{adar2008large}. 

To characterize the variability in time allocation across locations, we employ two metrics: the mean visitation frequency of each individual, $f_i= n_i/S_i$, where $n_i$ is the total time spent and $S_i$ is the number of unique locations visited by individual $i$, and the total dispersion in visitation across all locations, $\sigma_{f_i}$, allowing us to assess the degree of heterogeneity in how individuals allocate their time across different locations. We find that the visitation flux in the metaverse follows the scaling law $\sigma_{f} \propto \langle f\rangle^{\beta}$, where the exponent in the virtual world is $\beta_{vw} = 1.01$ (Fig. \ref{fig:individual_mobility} E), and in the network space it follows $\beta_{ethereum} = 1.01$ and $\beta_{polygon} = 0.98$ (Fig. \ref{fig:individual_mobility} F). The observed $\beta \sim 1$, indicates that as individuals explore more locations, the variability of their interests also grows linearly. 

We examined the location discovery rate of individuals, quantified as the number of unique locations visited, $S(n)$, after $n$ movements. We find that the exploration patterns of individuals follows $S(n) \propto n^{\mu}$, well-approximated by $\mu_{vw} = 0.52$ in the virtual world (Fig. \ref{fig:individual_mobility} G), and by $\mu_{ethereum} = 0.61$ and $\mu_{polygon} = 0.52$ in the contract space (Fig. \ref{fig:individual_mobility} H). This sub linear scaling suggests that as individuals move in the virtual space or purchase from more contracts, the tendency to visit new locations decreases. 

Finally, we ranked each location (land and contract) based on the number of times an individuals visited the location, so that $S^{*} = 3$ represents the third-most-visited location for the selected individual. We find that the probability of an individual returning to a specific location follows a power law, $P(S^{*}) \propto S^{-\alpha}$, characterized as $\alpha_{vw} = 1.1$ in the virtual world (Fig. \ref{fig:individual_mobility} I) and $\alpha_{ethereum} = 1.01$ and $\alpha_{polygon} = 1.8$ (Fig. \ref{fig:individual_mobility} J). That is, irrespective of the number of visited locations, individuals spend more than 50\% of their time in three to five locations and tend to return to previously visited locations 90\% of the time (Supplementary Figs. \ref{fig:si_time_spent_home}, \ref{fig:fpt_individual}), a pattern also characterizing mobility in the physical world\cite{gonzalez2008understanding}. 

\subsection*{Macroscopic patterns of mobility}

Despite corresponding to two separate systems within the metaverse - a virtual world with an imposed 2D structure and blockchain-based NFT contracts forming an infinite dimensional network— as Fig. \ref{fig:individual_mobility} illustrates, these systems exhibit a remarkable degree of quantitative similarity in individual mobility. This prompts us to ask, do the individual decisions driving these patterns are also driven by similar mechanisms, resulting in similar large-scale mobility patterns? To address this question, we develop a common network framework that captures mobility in both systems (Supplementary Section 4): (a) the virtual world network (VWN), where a node indicates a land (locations) and a link signifies temporal movement between the two lands. (b) the contract network (CN), where a node indicates a contract (location) and a link signifies temporal movement between the two contracts. 

We find that the degree distribution of nodes in both mobility networks have a heavy tail, where the probability of finding a location with degree $k$ scales as $P(k)\propto k^{-\alpha}$. In the virtual world network we find $\alpha_{vw} = 1.98$ (Fig. \ref{fig:meta-mobility-results} A), while in contract networks we find $\alpha_{ethereum} = 2.9$ and $\alpha_{polygon} = 2.4$ (Fig. \ref{fig:meta-mobility-results} B). The fact that $2 \lesssim \alpha<3 $ in Ethereum and Polygon contract networks, resulting in the divergence of $\langle k^2\rangle$, indicates that a few hub locations receive a disproportionate number of connections\cite{barabasi1999emergence, albert2002statistical, barrat2004architecture}. 


To assess movement between locations, we examine the number of individuals traveling between two locations, quantified via the link weight, $w_{ij}$. We find that in both mobility networks the link weight distributions follows a power law decay, $P(w_{ij}) \propto w_{ij}^{-\alpha}$, well-approximated by $\alpha_{vw} = 2.07$ in the virtual world network (Fig. \ref{fig:meta-mobility-results} D) and $\alpha_{ethereum} = 2.85$ and $\alpha_{polygon} = 2.5$ in the contract network (Fig. \ref{fig:meta-mobility-results} E). That is, if individuals moved randomly between locations, we would fail to observe such concentrated movement within a few specific locations (Supplementary Section 4.1). 

Finally, we find that a location's centrality within the network influences the number of individuals visiting that location: the number of visitors to a location follows $N_{S} \propto k^{\beta}$, where in the virtual world network we have $\beta_{vw} = 1.05$ (Fig. \ref{fig:meta-mobility-results} G), and in the contract network $\beta_{ethereum} = 1.05$ and $\beta_{polygon} = 1.13$ (Fig. \ref{fig:meta-mobility-results} H). This pattern remains consistent when examining the connection between centrality of the contract and the number of NFTs sold (Supplementary Fig. \ref{fig:si_deg_nft_sold}). The fact that $\beta \sim 1$ suggests that locations central within the meta-mobility network experience a linearly higher increase in the number of visitors compared to locations with fewer connections. 

\subsection*{Modeling the origins of metaverse mobility}

Our empirical results underscore four aspects of human mobility, each captured by a distinct scaling law: (1) individuals exhibit sub linear exploration patterns ($S(n) \propto n^{\mu}$ as $\mu<1$), (2) The frequency of visits to all locations follows a power law distribution ($P(S^{*}) \propto S^{-\alpha}$ as $1<\alpha<2$), and (3) the movement between the various locations defines a mobility network that exhibits a power law degree distribution ($P(k) \propto k^{-\alpha}$ with $2<\alpha<3$), and (4) the flow between locations (link weights) follows another power law distribution ($P(w_{ij}) \propto w_{ij}^{-\alpha}$ with $2<\alpha<3$). 


To understand how well the existing models can account for mobility in the metaverse, we start from the Exploration and Preferential Return (\textit{EPR}) model\cite{song2010modelling}, which has emerged as a foundational framework for understanding human mobility in the physical world\cite{ren2014predicting, yan2017universal, pappalardo2023future, schlapfer2021universal}. In the model, at each time step an individual decides to move to a randomly chosen new location based on the probability $p_{new} = \rho S^{-\gamma}$, where $S$ is the number of previously visited locations, or returns to a previously visited location based on their past visitation history with probability $1-p_{new}$ (Supplementary Section 5). Previous studies have estimated $\gamma_{EPR}=0.21$ in the physical world\cite{song2010modelling}, different from $\gamma_{vw} = 0.41 \pm 0.03$ we observe in the virtual world, and the values $\gamma_{ethereum} = 0.07 \pm 0.01$ and $\gamma_{polygon} = 0.18 \pm 0.01$ we observe in the contract space (Supplementary Fig. \ref{fig:model_specification_si} A-C). This implies that contract mobility, defined by $\gamma_{ethereum}$ and $\gamma_{polygon}$, experiences a slower decay rate in exploration compared to the physical world, showing a higher propensity to keep exploring new locations. On the other hand, the virtual world exhibits a faster decay rate (high $\gamma_{vw}$) compared to the physical world or the contract space, suggesting that individuals are less likely to explore new virtual locations\cite{eisenbeiss2012real}. 

The \textit{EPR} model predicts the individual visitation frequency of locations, $P(S^{*})\propto S^{-\alpha_{EPR}}$, with exponent $\alpha_{EPR} = 1.42 \pm 0.03$, together with a sub-linear exploration patterns in new visitation, $S \propto n^{\mu_{EPR}}$, as $\mu_{EPR} = 0.7 \pm 0.01$ (Table \ref{tab:model_exponents_si}, Supplementary Section 5.4). This shows that the \textit{EPR} model is able to account for the observed sub-linear visitation patterns and the power law distribution in visited locations ($\alpha_{vw} = 1.35 \pm 0.03$, $\mu_{vw} = 0.52 \pm 0.01$), as highlighted in key observations (1) and (2). 

\begin{figure}[ht]
\centering
\includegraphics[width=\linewidth, height = 28em, keepaspectratio]{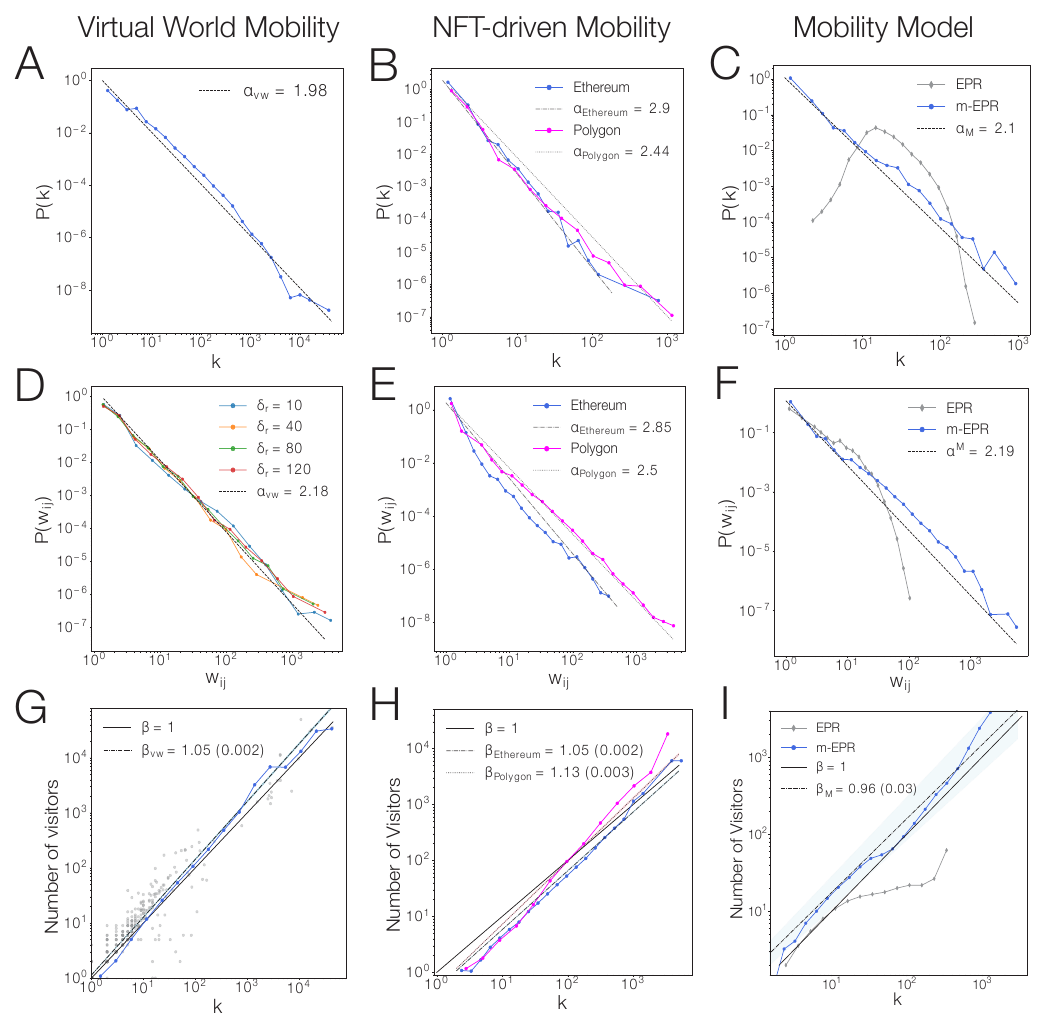}
\caption{\textbf{Capturing models of meta-mobility.} 
\textbf{(A)} Degree distribution of lands in the mobility network. We observe a fat tailed distribution in degrees, well-approximated by a power law, $P(k) \propto k^{-\alpha}$ as $\alpha_{vw} = 1.98$. 
\textbf{(B)} Degree distribution of contracts in the contract network. We observe a heavy tailed distribution where few contracts receive most of the connections, well-approximated by a power law, $P(k) \propto k^{-\alpha}$, where $\alpha_{ethereum} = 2.9$ and $\alpha_{polygon} = 2.4$. 
\textbf{(C)} Degree distribution from the model simulations. We show the results from the EPR model and the m-EPR model, finding that the m-EPR model recreates the heterogeneous degree distribution as $\alpha^{M} = 2.1 \pm 0.06$. 
We show the link weight, $w_{ij}$, distribution of the network for \textbf{(D)} virtual world mobility and \textbf{(E)} contract mobility. The link weight distribution follows a power law decay as $P(w_{j}) \propto w_{ij}^{-\alpha}$, where $\alpha_{vw}= 2.18$ in the virtual world and $\alpha_{ethereum} = 2.85$; $\alpha_{polygon} = 2.5$ in the contract network. We display link weight distributions between nodes at different physical distances ($\delta_r$). 
\textbf{(F)} Link weight distribution from model simulations. We find that the m-EPR model is able to uncover the concentrated flows between locations with $\alpha^{M} =2.19 \pm 0.03$ in the network. \textbf{(G)} Relationship between degree of land in the virtual world mobility network and its number of visitors. We find that the two variables scales as $N_{S} \propto k^{\beta}$, where $\beta_{vw} = 1.05$. \textbf{(H)} Relationship between degree of contract in the contract mobility network and its number of visitors. We find that $N \propto k^{\beta}$, where $\beta_{ethereum} = 1.05$ and $\beta_{polygon} = 1.13$. 
\textbf{(I)} Relationship between degree and visitors from model simulations. We find that the\textit{m-EPR} model obtains a positive scaling between the two variables ($\beta^{M} = 0.96 \pm 0.03$).}
\label{fig:meta-mobility-results}
\end{figure}

Yet, as the agents select locations randomly, the \textit{EPR} model is unable to uncover the emergence of visitation hubs and the network-based relationship between the locations as documented in Fig. \ref{fig:meta-mobility-results}, i.e., the patterns (3) and (4). In particular, the network structure predicted by the \textit{EPR} model is not scale-free (Fig. \ref{fig:meta-mobility-results} C), violating observation (3). It also fails to predict heterogeneous visitation patterns, i.e. observation (4), finding link weights that are several magnitudes lower than empirical data, and fails to capture the linear correlation between hubs and their visitation numbers (Fig. \ref{fig:meta-mobility-results} F, I). In fact, the mobility networks generated by the EPR model closely resembles random movements without preferential return (Supplementary Fig. \ref{fig:si_epr_net_dist}), suggesting that the model offers inadequate explanations about collective mobility, as highlighted in observations (3) and (4).

\begin{figure}[ht]
\centering
\includegraphics[width=25em, height = 18em, keepaspectratio]{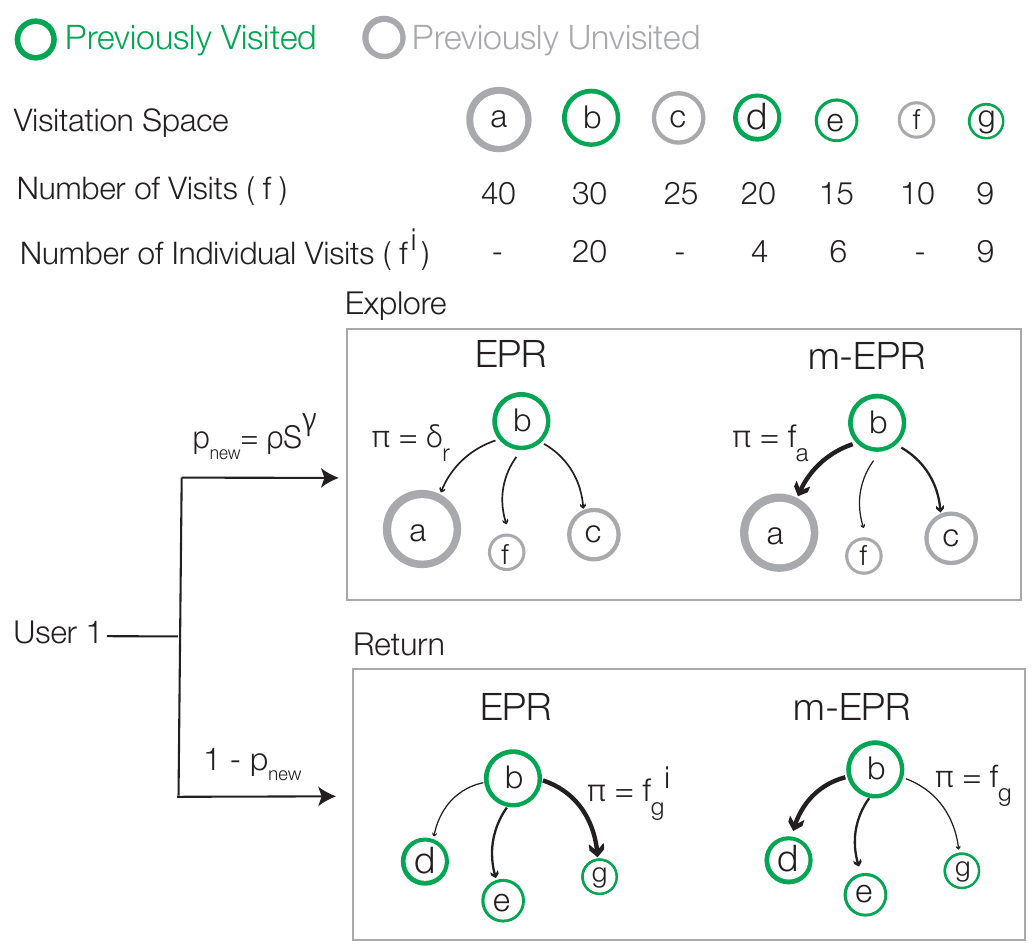}
\caption{\textbf{Foundation of the mobility model.} We first organize the possible visiting locations $a$ to $g$, into whether the location has been previously visited (green) or unvisited (gray). At each movement, an individual decides to explore a new location (gray) with probability $p_{new} \propto S^{-\gamma}$, where $S$ represents the number of previously visited locations. With its complementary probability, $1-p_{new}$, an individual decides to revisit a previously visited location (green). In the EPR formulation, when visiting a new location, the individual randomly selects a new location drawn at some distance $\delta_r$. In the proposed \textit{m-EPR} model, the individual is biased towards the more popular locations, sampled according to the probability $\pi = m_a/\sum_{j} m_j$, where $m_a$ is the popularity of the location $a$. When deciding to revisit a previously visited location, in the \textit{EPR} model the individual selects based on their individual past visitation history, $\pi_i = m^{i}_g/\sum_{j} m^{i}_j$, where $m^{i}_g$ represents the number of visits to the location $g$ by individual $i$. In contrast, in the \textit{m-EPR} model, an individual revisits a location according to the probability $\pi = m_g/\sum_{j} m_j$, where $m_g$ represents the total number of visits to the location $i$ by all individuals. In contrast to the EPR model, where an individual randomly selects a new location and preferentially re-visits location based on only their individual visitation history ($\pi_i$), the \textit{m-EPR} model allows individuals to select new locations and revisit old locations based on their global popularity ($\pi$).}
\label{fig:model_schema}
\end{figure}

To resolve this discrepancy, we extend the \textit{EPR} model by incorporating the relative popularity of each location, and biasing the individual movements towards more popular locations (Supplementary Section 5.3). In this new metaverse-adopted EPR model, that we call the \textit{m-EPR} model (Fig. \ref{fig:model_schema}), an individual decides to move to a new location with probability $p_{new} = \rho S^{-\gamma}$, or return to a previously visited location with probability $1-p_{new}$, mirroring the \textit{EPR} model. Yet, in both scenarios we add a new element: before making a move, an individual evaluates the popularity of all available locations, and moves according to the transition probability, $\pi_j =  m_j/\sum_i m_i$, where $m_j$ is the number of visits to location $j$ by all agents. Unlike the \textit{EPR} model and its several proposed variants\cite{ren2014predicting, yan2017universal, pappalardo2023future, moro2021mobility, schlapfer2021universal}, the \textit{m-EPR} model informs the individual movement based on the location's visitation numbers, normalized across all locations irrespective of distance. Indeed, the empirical data indicates that an individual is significantly more inclined to visit a location with higher visitation numbers compared to those with fewer visitation numbers ($\beta = 0.95$, $R^{2} = 0.88$), offering the empirical rationale for the added component (Supplementary Fig. \ref{fig:model_specification_si} D-F). 

To explore the predictive power of the proposed \textit{m-EPR} model, we performed simulations using parameters derived from the virtual world ($\gamma_{vw} = 0.41$), and explored the model-based predictions by determining the scaling exponents related to individual mobility and the mobility network (Table \ref{tab:model_exponents_si}). We also examined simulations with different temporal and spatial parameters (Supplementary Figs. \ref{fig:si_mepr_stat_cond}, \ref{fig:si_mepr_n_locs_cond}). In each case, we find a remarkable alignment with the empirical observations. 


To be specific, we find that the \textit{m-EPR} model recreates the individual mobility patterns (key observations (1) and (2)), by predicting a sub linear behavior in exploration, $S \propto n^{\mu_M}$ with $\mu_{M} = 0.6 \pm 0.004$, closely related to the empirically observed values of $\mu_{vw}= 0.52 \pm 0.01$ (Supplementary Fig. \ref{fig:si_mepr_ind_mob} A). The model also identifies a power law visitation frequency distribution, $P(S^{*}) \propto S^{-\alpha_{M}}$, where $\alpha_{M} = 1.41 \pm 0.02$, consistent with empirical observations of $\alpha_{vw}= 1.35 \pm 0.03$ (Supplementary Fig. \ref{fig:si_mepr_ind_mob} B). 

Importantly, the \textit{m-EPR} model can successfully capture the emergent characteristics of the observed mobility network (key observations (3) and (4)). Specifically, we observe that the \textit{m-EPR} model predicts a scale-free network with degree distribution, $P(k) \propto k^{-\alpha_{M}}$, with $\alpha_{M} = 2.1\pm 0.06$ (Fig. \ref{fig:meta-mobility-results} C), closely aligned with empirical observations ($\alpha_{vw} = 1.98 \pm 0.01$). Further, the \textit{m-EPR} model accurately captures the link weight distribution, $P(w_{ij}) \propto w_{ij}^{\alpha_{M}}$, with exponent $\alpha_{M} = 2.19 \pm 0.03$ (Fig. \ref{fig:meta-mobility-results} F), suggesting that the model explains the visitation heterogeneity that emerges in the mobility network ($\alpha_{vw} = 2.18 \pm 0.08$). Finally, the \textit{m-EPR} model, by encouraging individual movements to popular locations, successfully captures the positive relationship between a location's degree and the number of visitors, $N(S) \propto k^{\beta_{M}}$, where $\beta_{M} = 0.96 \pm 0.03$ (Fig. \ref{fig:meta-mobility-results} I), consistent with $\beta_{vw} = 1.05 \pm 0.002$) found in the empirical data. 

Indeed, the \textit{EPR} model, which assumes that individuals select their destinations independently of each other, overlooks the interconnected dynamics of exploration together with the fact that some mobility patterns are driven by the previous popularity of the visited location. It thus fails to uncover the heterogeneous flows between locations. In contrast, the \textit{m-EPR} predicts mobility network and heterogeneous mobility flows, hence capturing the essential features of the collective mobility patterns in the metaverse. The ability of the \textit{m-EPR} model to accurately describe the aggregate level insights of the mobility network implies that individual movement in the metaverse is driven by popularity based dynamics, a feature absent from human mobility within the physical space. 

\section*{Discussion}

The metaverse offers a freedom of mobility, releasing individuals from the traditional geographical constraints and limitations imposed by commuting times and costs\cite{bainbridge2007scientific}. In theory, this newfound freedom of movement offers the potential for a more equitable distribution of access and human activity patterns\cite{mazalek2011embodying}. Yet, we find that the exploration patterns of individuals, in both movements in the virtual world and purchasing of new NFTs, results in highly uneven distribution of visitations across different locations. These empirical findings prompted us to enrich the existing human mobility models, adapting them to the metaverse environment. The resulting \textit{m-EPR} model leverages a popularity-driven mechanism to successfully replicate individual level characteristics and the aggregate mobility patterns in both the virtual world and the contract space. The model's ability to explain the observed quantitative patterns is rooted in a simple prospective mechanism: an individual's inclination to visit a specific location is influenced by the number of previous visitors to that location, irrespective of his/her own visitation history or distances between locations. This mechanism, absent from mobility in the physical world, fundamentally alters the quantitative patterns of metaverse mobility. 

At the same time, individual movements in the metaverse preserves multiple properties observed in the physical world. For example, an individual's inclination to explore new locations diminishes over time, and individuals display substantial heterogeneity in visitation preference across all visited locations. These consistent patterns underscore the persistent elements of human behavior in mobility, holding relevance in both physical and metaverse environments. 


Note that virtual worlds also contain a three-dimensional component\cite{dionisio20133d}. For example, individuals could hike a mountain or hang out at the top of a building, dimensions captured by the data we collected and available for future work. The current modeling framework overlooks these extra dimensions, along with other features that influence human decisions in mobility, such as the impact of social networks on location choices, the different utility or aesthetic appeal of the available locations, and the potential activities within specific points of interest— such as playing poker at a virtual casino\cite{bloomberg}, or viewing digital art at Sotheby's virtual gallery\cite{decentraland}. Future work should address their role, and may also design controlled experiments to understand the value of autonomous generative agents in enabling social interactions and location discovery\cite{park2023generative, zhu2023ghost, kaiya2023lyfe, liu2023training}. 

\section*{Methods}

\subsection*{Constructing meta-mobility systems}

\begin{figure}[ht]
\centering
\includegraphics[width=\linewidth, height = 30em, keepaspectratio]{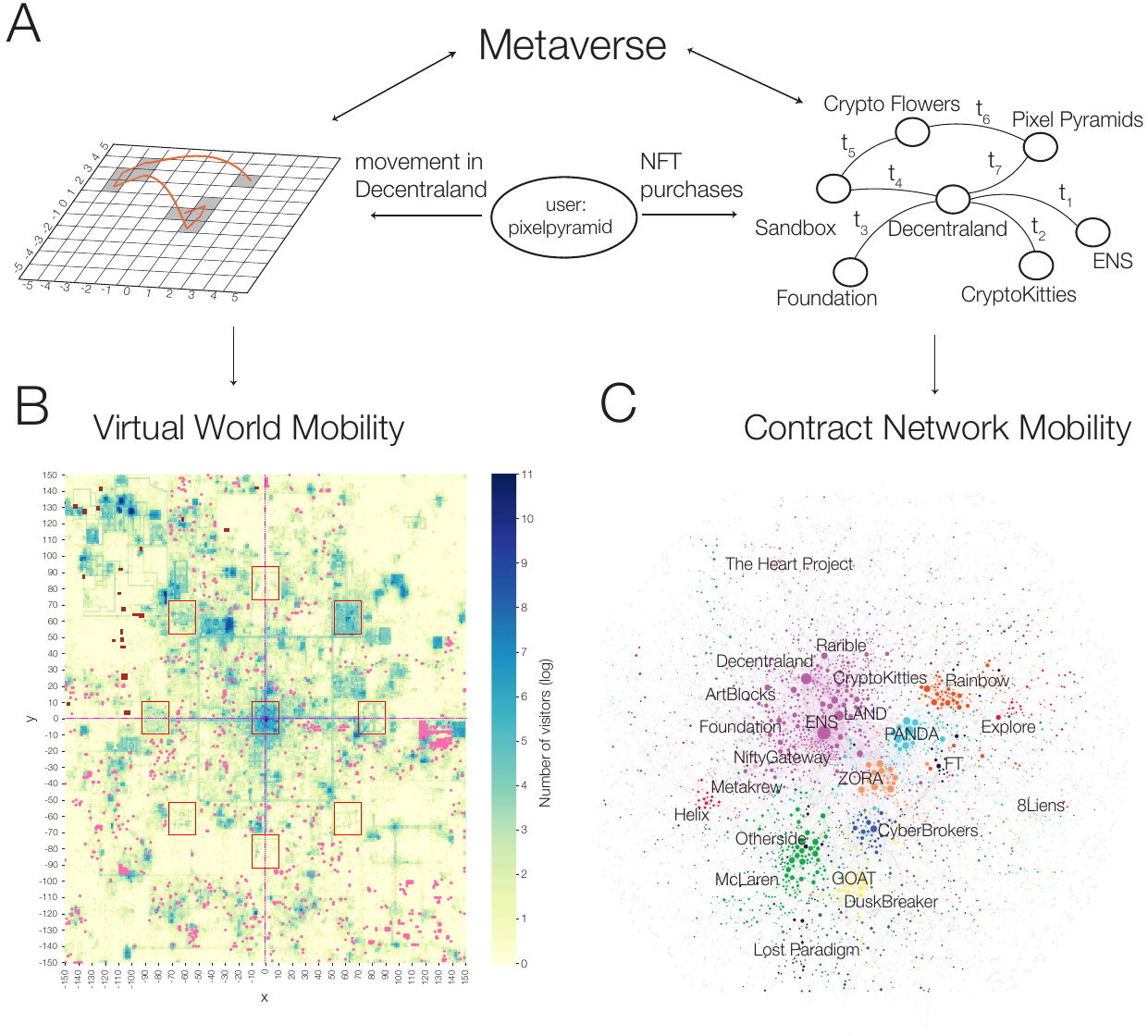}
\caption{\textbf{Measuring meta-mobility.} 
\textbf{(A)} To quantify mobility in the metaverse, we explore two separate datasets: Mobility in a 2-dimensional virtual world (left) and in the network space (right). In the virtual world, an individual, posing as an avatar, can move through a 2D virtual environment via local movements or teleportation. Using the trajectories of all individuals in the virtual world, we created a mobility network, whose nodes are lands and a link captures the movement between two locations. We track the mobility of the same individuals in the network space, capturing virtual marketplaces through their NFT purchases, allowing us to build a time-resolved contract network. For example, a user with screen name \textit{pikelpyramid}, purchased an NFT from the \textit{Decentraland} contract and the subsequently purchased a new NFT from \textit{Sandbox}, creating a link between \textit{Decentraland} and \textit{Sandbox}. 
\textbf{(B)} Visualization of the Decentraland virtual world. Each land in the virtual world is identified by its (x, y) coordinates, organized in a 2D symmetrical layout. The lands are colored based on the number of visitors. We mark the parcels that were sold during our observation period, with points sized based on the selling price of land.
\textbf{(C)} The Ethereum contract network traveled by our users. The nodes (contracts) are sized based on the number of users that visit the contract, and the top 10 network communities are colored for clarity.}
\label{fig:global_mobility}
\end{figure}

\textbf{Virtual world mobility.} The data regarding the movements of individuals is derived from Decentraland, one of the first decentralized virtual worlds. We conduct two data collection processes (Supplementary Section 1). First data collection process lasted from March 15, 2022 to Aug 6, 2022, extracting data from a single data server, resulting in 81,563 users and 110,416,682 displacements (D1), and the second data collection lasted from Aug 7,2022 to September 19, 2022, capturing 141,226,580 movements by 94,149 users and (D2). Finally, we collect 15,209 sales of 6,773 lands, and 1,562 sales of 1,075 estates (collection of land) comprising of 7,159 lands. In the mobility network, a node is a land and a link signifies movement between the two locations. We create the mobility network by aggregating the movements of all individuals (Fig. \ref{fig:global_mobility} A, Supplementary Section 1.1). 

\textbf{Contract network mobility.} Each Non-Fungible Token (NFT) is associated with a specific contract, either of type ERC-721 or ERC-1155, representing transaction rules and royalty rates for each NFT transfer. An individual follows the rules set by the contract to purchase an NFT, similar to how collectors purchase an art item (NFT) from an art gallery (contract). For the same set of individuals, whose wallets we followed on Decentraland, we extract data regarding NFT purchases from two different blockchains: (a) Ethereum, finding 1,165,310 NFTs from 23,827 contracts collected by 14,732 (9\%) users, and (b) Polygon, find 3,112,300 NFTs from 54,918 contracts by 41,870 (25\%) users. Using the NFT purchase history of each individual, we build a contract network, where a NFT contract for \textit{Foundation} will be connected to another NFT contract for \textit{Artblocks}, if an individual purchased a NFT from \textit{Foundation} and then subsequently purchased an NFT from \textit{Artblocks} (Fig. \ref{fig:global_mobility} A). The resulting contract network captures similarities between contracts and also characterises mobility in the NFT space (Fig. \ref{fig:global_mobility} C, Supplementary Section 1.3).

\subsection*{Modelling metaverse mobility}

\textbf{EPR model.} The simulations are conducted on a 300x300 location grid, and the choice of the location is sampled from the jump distribution derived from empirical data. The probability that an individual chooses to explore a new location is given by $p_{new} \propto S^{-\gamma}$, where S is the number of locations visited by the individual so far. With probability, $1-p_{new}$, the individual decides to revisit a previously visited location, proportional to the number of past visits. 

\textbf{m-EPR model.} The individual decides to explore a new location with probability $p_{new} \propto S^{-\gamma}$, and with probability $1-p_{new}$, the revisit a previously visited location. In contrast to the \textit{EPR} model, the individual movements are influenced by the popularity of each location, characterized by the number of visits to that location, $\pi = m_g/\sum_{j} m_j$, where $m_g$ represents the total number of visits to the location $i$ by all individuals. To do this, we keep track of the number of visitors, $m_t (i)$ to a location $i$. At the end of each time step, the number of visitors to the location, $m_{t+1} {i}$ is updated. 

\textbf{Simulation strategy.} The simulations were conducted for $n=5,000$ agents and $S=20,000$ locations in a infinite dimensional space, meaning the distance between locations does not affect the transition probability. At time $t=0$, the $n$ agents are randomly distributed to $S$ locations. At each time step, an agent becomes active proportional to their activity level drawn from the empirical distribution, and make $m=4$ movements (the number of steps is arbitrary and does not influence the macroscopic patterns). The agent visits a new location with probability $p_{new}$ and with probability $1-p_{new}$ they revisit a previously visited location. The choice of the location is decided based on the number of visits by all agents. The individual trajectories were simulated until the visitation patterns achieved a stationary condition (see Supplementary Section 5). The simulations followed discrete time movement, and it does not incur any waiting time prior to each movement. 

\section*{Data availability}

The dataset used in this study along with the codes required to get the latest data is provided at \url{https://github.com/Barabasi-Lab/metaverse-mobility}.


\bibliography{scibib}

\begin{thebibliography}{10}

\bibitem{caldarelli2023role}
G.~Caldarelli, {\it et~al.\/}, {\it Nature Computational Science\/} pp. 1--8 (2023).

\bibitem{ball2022metaverse}
M.~Ball, {\it The metaverse: and how it will revolutionize everything\/} (Liveright Publishing, 2022).

\bibitem{mkinsey}
Mckinsey, Value creation in the metaverse, \url{https://www.mckinsey.com/capabilities/growth-marketing-and-sales/our-insights/value-creation-in-the-metaverse} (2022).

\bibitem{gadekallu2022blockchain}
T.~R. Gadekallu, {\it et~al.\/}, {\it arXiv preprint arXiv:2203.09738\/}  (2022).

\bibitem{marthews2019blockchain}
A.~Marthews, C.~E. Tucker, {\it Cryptoassets: Legal and Monetary Perspectives\/}  (2019).

\bibitem{weyl2022decentralized}
E.~G. Weyl, P.~Ohlhaver, V.~Buterin, {\it Available at SSRN 4105763\/}  (2022).

\bibitem{vasan2022quantifying}
K.~Vasan, M.~Janosov, A.-L. Barab{\'a}si, {\it Scientific reports\/} {\bf 12}, 1 (2022).

\bibitem{nadini2021mapping}
M.~Nadini, {\it et~al.\/}, {\it Scientific reports\/} {\bf 11}, 20902 (2021).

\bibitem{gomez2023promise}
D.~G{\'o}mez-Zar{\'a}, P.~Schiffer, D.~Wang, {\it Nature Human Behaviour\/} pp. 1--4 (2023).

\bibitem{gonzalez2008understanding}
M.~C. Gonzalez, C.~A. Hidalgo, A.-L. Barabasi, {\it nature\/} {\bf 453}, 779 (2008).

\bibitem{yuan2012discovering}
J.~Yuan, Y.~Zheng, X.~Xie, {\it Proceedings of the 18th ACM SIGKDD international conference on Knowledge discovery and data mining\/} (2012), pp. 186--194.

\bibitem{zhao2021characteristics}
C.~Zhao, A.~Zeng, C.~H. Yeung, {\it EPJ Data Science\/} {\bf 10}, 5 (2021).

\bibitem{simini2012universal}
F.~Simini, M.~C. Gonz{\'a}lez, A.~Maritan, A.-L. Barab{\'a}si, {\it Nature\/} {\bf 484}, 96 (2012).

\bibitem{crandall2010inferring}
D.~J. Crandall, {\it et~al.\/}, {\it Proceedings of the National Academy of Sciences\/} {\bf 107}, 22436 (2010).

\bibitem{eagle2010network}
N.~Eagle, M.~Macy, R.~Claxton, {\it Science\/} {\bf 328}, 1029 (2010).

\bibitem{davis2009avatars}
A.~Davis, J.~Murphy, D.~Owens, D.~Khazanchi, I.~Zigurs, {\it Journal of the Association for Information Systems\/} {\bf 10}, 1 (2009).

\bibitem{schelling1971dynamic}
T.~C. Schelling, {\it Journal of mathematical sociology\/} {\bf 1}, 143 (1971).

\bibitem{echenique2007measure}
F.~Echenique, R.~G. Fryer~Jr, {\it The Quarterly Journal of Economics\/} {\bf 122}, 441 (2007).

\bibitem{netto2015segregated}
V.~M. Netto, M.~P. Soares, R.~Paschoalino, {\it International Journal of Urban and Regional Research\/} {\bf 39}, 1084 (2015).

\bibitem{moro2021mobility}
E.~Moro, D.~Calacci, X.~Dong, A.~Pentland, {\it Nature communications\/} {\bf 12}, 4633 (2021).

\bibitem{napoli2023socioeconomic}
L.~Napoli, V.~Sekara, M.~Garc{\'\i}a-Herranz, M.~Karsai, {\it Proceedings of the National Academy of Sciences\/} {\bf 120}, e2305285120 (2023).

\bibitem{bonaccorsi2020economic}
G.~Bonaccorsi, {\it et~al.\/}, {\it Proceedings of the National Academy of Sciences\/} {\bf 117}, 15530 (2020).

\bibitem{kojaku2021effectiveness}
S.~Kojaku, L.~H{\'e}bert-Dufresne, E.~Mones, S.~Lehmann, Y.-Y. Ahn, {\it Nature physics\/} {\bf 17}, 652 (2021).

\bibitem{chang2021mobility}
S.~Chang, {\it et~al.\/}, {\it Nature\/} {\bf 589}, 82 (2021).

\bibitem{santana2023covid}
C.~Santana, {\it et~al.\/}, {\it Nature Human Behaviour\/} {\bf 7}, 1729 (2023).

\bibitem{szell2012understanding}
M.~Szell, R.~Sinatra, G.~Petri, S.~Thurner, V.~Latora, {\it Scientific reports\/} {\bf 2}, 1 (2012).

\bibitem{hu2018life}
T.~Hu, J.~Luo, W.~Liu, {\it Proceedings of the International AAAI Conference on Web and Social Media\/} (2018), vol.~12.

\bibitem{hazarie2020uncovering}
S.~Hazarie, H.~Barbosa, A.~Frank, R.~Menezes, G.~Ghoshal, {\it Journal of the Royal Society Interface\/} {\bf 17}, 20200250 (2020).

\bibitem{zhong2017revealing}
C.~Zhong, {\it et~al.\/}, {\it Urban Studies\/} {\bf 54}, 437 (2017).

\bibitem{yan2017universal}
X.-Y. Yan, W.-X. Wang, Z.-Y. Gao, Y.-C. Lai, {\it Nature communications\/} {\bf 8}, 1639 (2017).

\bibitem{radicchi2012rationality}
F.~Radicchi, A.~Baronchelli, L.~A. Amaral, {\it PloS one\/} {\bf 7}, e29910 (2012).

\bibitem{bainbridge2007scientific}
W.~S. Bainbridge, {\it science\/} {\bf 317}, 472 (2007).

\bibitem{bertaud2021order}
A.~Bertaud, {\it Town and Regional Planning\/} {\bf 79}, 2 (2021).

\bibitem{guimera2005worldwide}
R.~Guimera, S.~Mossa, A.~Turtschi, L.~N. Amaral, {\it Proceedings of the National Academy of Sciences\/} {\bf 102}, 7794 (2005).

\bibitem{west2018scale}
G.~West, {\it Scale: The universal laws of life, growth, and death in organisms, cities, and companies\/} (Penguin, 2018).

\bibitem{hasan2013understanding}
S.~Hasan, X.~Zhan, S.~V. Ukkusuri, {\it Proceedings of the 2nd ACM SIGKDD international workshop on urban computing\/} (2013), pp. 1--8.

\bibitem{zukin2009new}
S.~Zukin, {\it et~al.\/}, {\it City \& Community\/} {\bf 8}, 47 (2009).

\bibitem{hong2020universal}
I.~Hong, M.~R. Frank, I.~Rahwan, W.-S. Jung, H.~Youn, {\it Science advances\/} {\bf 6}, eaba4934 (2020).

\bibitem{alessandretti2020scales}
L.~Alessandretti, U.~Aslak, S.~Lehmann, {\it Nature\/} {\bf 587}, 402 (2020).

\bibitem{barbosa2018human}
H.~Barbosa, {\it et~al.\/}, {\it Physics Reports\/} {\bf 734}, 1 (2018).

\bibitem{bokanyi2021universal}
E.~Bok{\'a}nyi, S.~Juh{\'a}sz, M.~Karsai, B.~Lengyel, {\it Scientific reports\/} {\bf 11}, 1 (2021).

\bibitem{schlapfer2021universal}
M.~Schl{\"a}pfer, {\it et~al.\/}, {\it Nature\/} {\bf 593}, 522 (2021).

\bibitem{adar2008large}
E.~Adar, J.~Teevan, S.~T. Dumais, {\it Proceedings of the SIGCHI conference on Human Factors in Computing Systems\/} (2008), pp. 1197--1206.

\bibitem{barabasi1999emergence}
A.-L. Barab{\'a}si, R.~Albert, {\it science\/} {\bf 286}, 509 (1999).

\bibitem{albert2002statistical}
R.~Albert, A.-L. Barab{\'a}si, {\it Reviews of modern physics\/} {\bf 74}, 47 (2002).

\bibitem{barrat2004architecture}
A.~Barrat, M.~Barthelemy, R.~Pastor-Satorras, A.~Vespignani, {\it Proceedings of the national academy of sciences\/} {\bf 101}, 3747 (2004).

\bibitem{song2010modelling}
C.~Song, T.~Koren, P.~Wang, A.-L. Barab{\'a}si, {\it Nature physics\/} {\bf 6}, 818 (2010).

\bibitem{ren2014predicting}
Y.~Ren, M.~Ercsey-Ravasz, P.~Wang, M.~C. Gonz{\'a}lez, Z.~Toroczkai, {\it Nature communications\/} {\bf 5}, 5347 (2014).

\bibitem{pappalardo2023future}
L.~Pappalardo, E.~Manley, V.~Sekara, L.~Alessandretti, {\it Nature Computational Science\/} pp. 1--13 (2023).

\bibitem{eisenbeiss2012real}
M.~Eisenbeiss, B.~Blechschmidt, K.~Backhaus, P.~A. Freund, {\it Journal of Interactive Marketing\/} {\bf 26}, 4 (2012).

\bibitem{mazalek2011embodying}
A.~Mazalek, S.~Chandrasekharan, M.~Nitsche, T.~Welsh, P.~Clifton, {\it Reinventing ourselves: contemporary concepts of identity in virtual worlds\/} (Springer, 2011), pp. 129--151.

\bibitem{dionisio20133d}
J.~D.~N. Dionisio, W.~G.~B. Iii, R.~Gilbert, {\it ACM Computing Surveys (CSUR)\/} {\bf 45}, 1 (2013).

\bibitem{bloomberg}
C.~D'Anastasio, Sotheby’s opens a virtual gallery in decentraland, \url{https://www.bloomberg.com/news/articles/2022-03-01/what-is-decentraland-metaverse-is-often-used-for-crypto-poker} (2022).

\bibitem{decentraland}
Decentraland, Sotheby’s opens a virtual gallery in decentraland, \url{https://decentraland.org/blog/announcements/sotheby-s-opens-a-virtual-gallery-in-decentraland} (2021).

\bibitem{park2023generative}
J.~S. Park, {\it et~al.\/}, {\it Proceedings of the 36th Annual ACM Symposium on User Interface Software and Technology\/} (2023), pp. 1--22.

\bibitem{zhu2023ghost}
X.~Zhu, {\it et~al.\/}, {\it arXiv preprint arXiv:2305.17144\/}  (2023).

\bibitem{kaiya2023lyfe}
Z.~Kaiya, {\it et~al.\/}, {\it arXiv preprint arXiv:2310.02172\/}  (2023).

\bibitem{liu2023training}
R.~Liu, {\it et~al.\/}, {\it arXiv preprint arXiv:2305.16960\/}  (2023).

\bibitem{west2018scale_si}
G.~West, {\it Scale: The universal laws of life, growth, and death in organisms, cities, and companies\/} (Penguin, 2018).

\bibitem{newman2018networks}
M.~Newman, {\it Networks\/} (Oxford university press, 2018).

\end{thebibliography}
\bibliographystyle{Science}

\section*{Acknowledgments}

This work was supported by the Templeton Foundation under contract 61066, the Air Force Office of Scientific Research under award number FA9550-19-1-0354, and the Eric and Wendy Schmidt Fund for Strategic Innovation (G-22-63228), and the National Science Foundation (SES-2219575). ALB was also supported by European Research Council (ERC) Synergy grant (DYNASNET-810115). M.K. was supported by the Accelnet-Multinet NSF grant. M.K. acknowledges funding from the National Laboratory for Health Security (RRF-2.3.1-21-2022-00006); the SoBigData++ H2020-871042 project and the MOMA WWTF project. 

We thank the data providers of this study: Etherscan, a block explorer platform for Ethereum; Polygonscan, a block explorer for Polygon, and the Decentraland Foundation. The project was granted Institutional Review Board (IRB) exemption by Northeastern University (Reference number 23-03-29). 

\clearpage

%
\setcounter{figure}{0}  
\setcounter{table}{0}  

\renewcommand{\thefigure}{S\arabic{figure}}

\renewcommand{\thetable}{S\arabic{table}}


\topmargin 0.0cm
\oddsidemargin 0.2cm
\textwidth 16cm 
\textheight 21cm
\footskip 1.0cm

\baselineskip24pt
\section*{Supplementary Information}

\section{Data collection and processing}  
\subsection{Virtual world mobility}

Decentraland (https://decentraland.org/) is a virtual world that exists on a 2-dimensional (2D) grid. The software to create, run and deploy the metaverse is publicly accessible on Github website (https://github.com/decentraland), and authorized servers use the code to render 3D objects, contents, and enable communication between users. The server architecture allows an inter-connected system where each server is required to post the locations of the users to dynamically update communication nodes between users (https://docs.decentraland.org). 

\begin{figure}[!ht]
\centering
\includegraphics[width=\linewidth, height = 20em, keepaspectratio]{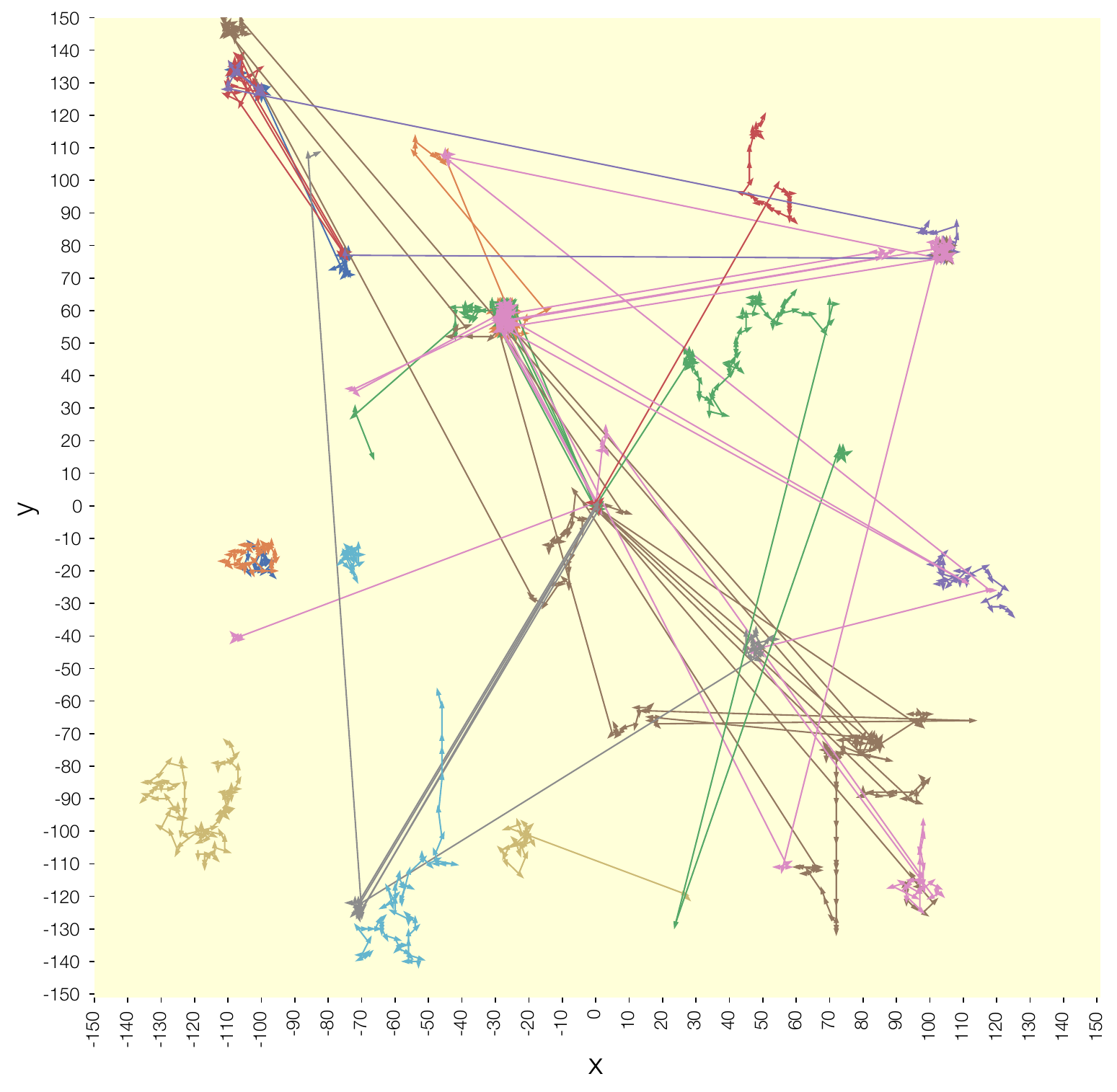}
\caption{\textbf{Individual mobility in the virtual space.} We show examples of mobility in the virtual space by 20 random individuals in the metaverse. These trajectories highlight the use of large jump distances followed by small movements in nearby locations.}
\label{fig:ind_mobility_example}
\end{figure}

\begin{figure}[ht]
\centering
\includegraphics[width=\linewidth, keepaspectratio]{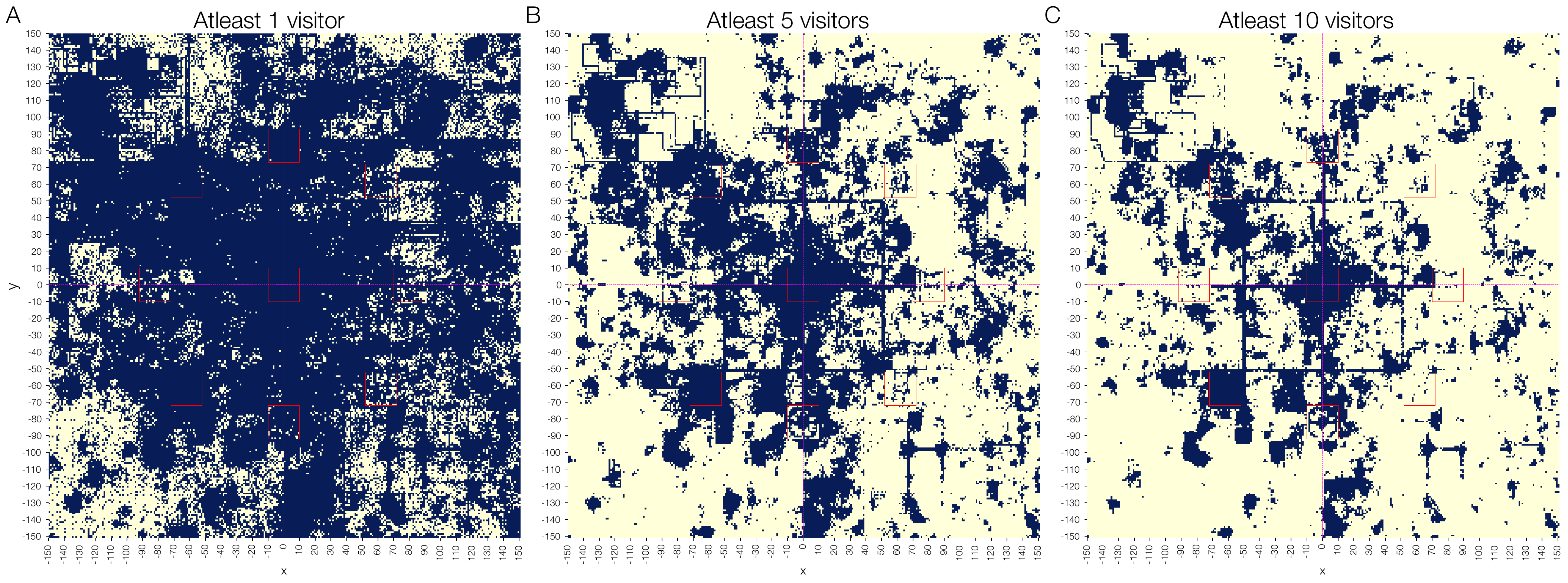}
\caption{\textbf{Locations explored at different subsets of visitation numbers.} We highlight the locations hat have had at least \textbf{(A)} 1 visitor \textbf{(B)} 5 visitors and \textbf{(C)} 10 visitors. This shows that most of the parcels struggle to attract new visitors. }
\label{fig:si_subset_exploration}
\end{figure}

\begin{figure}[ht]
\centering
\includegraphics[width=\linewidth, height = 20em, keepaspectratio]{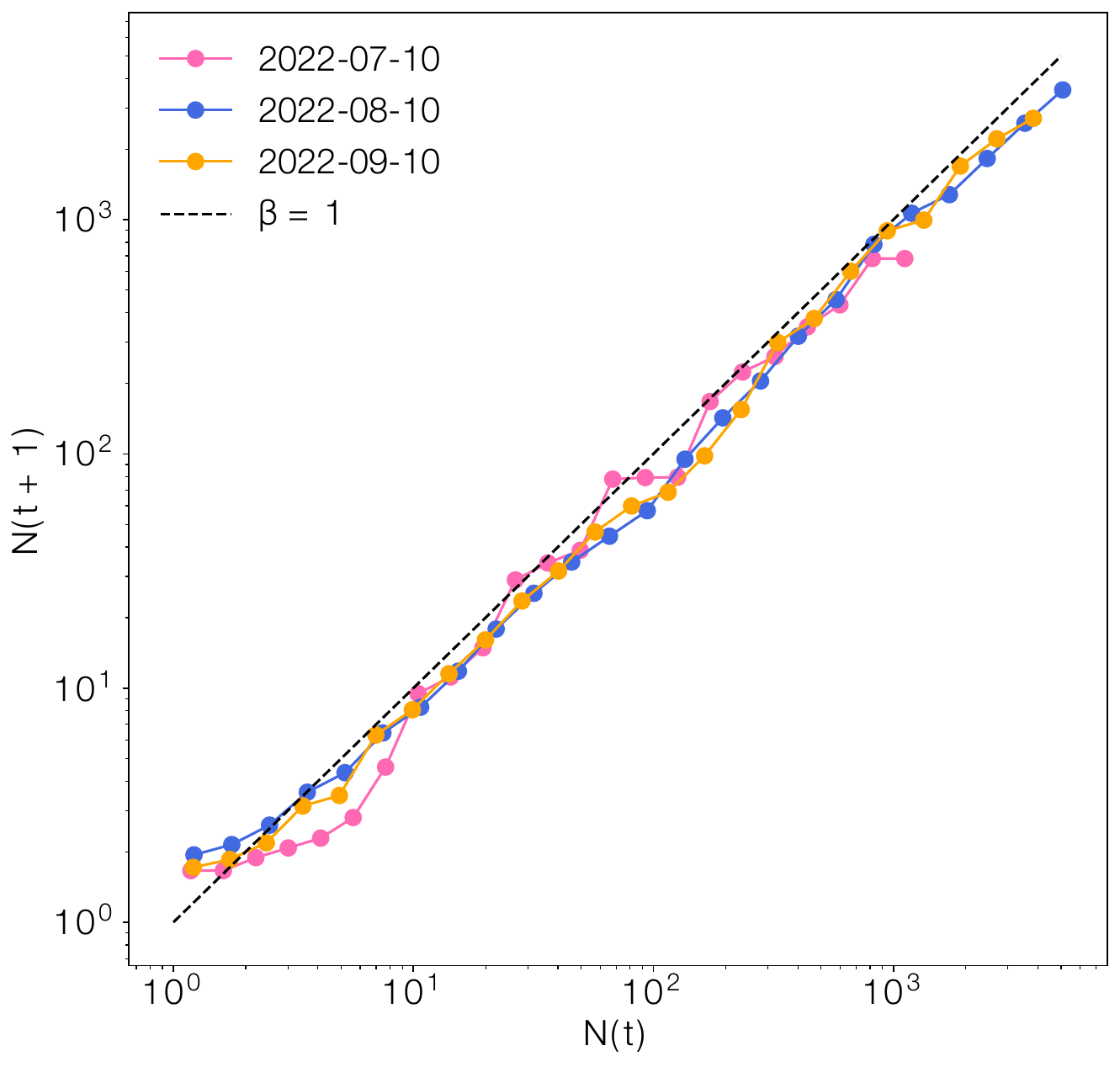}
\caption{\textbf{Trends in visitation numbers.} We find that the number of visitors a location receives at the next time step \textit{t+1} is similar to the number of visitors at the current time step \textit{t}. This indicates a stability in visitation patterns across locations.}
\label{fig:si_pa_evidence}
\end{figure}

We created a bot that automatically pings these servers every ten seconds to extract the precise location of each individual on the platform (Supplementary Fig. \ref{fig:ind_mobility_example}). The first data collection process lasted from March 15, 2022 to Aug 6, 2022, crawling data from a single server, resulting in data about 81,563 users and 110,416,682 displacements (D1). To ensure that the data is not affected by irregularities pertinent to a single server, we collect additional data from multiple data servers from Aug 7,2022 to September 19, 2022, resulting in 94,149 users and 141,226,580 movements (D2). The combined data captures the mobility of 163,770 users and their 251,643,262 movements across locations in the virtual world (Supplementary Fig. \ref{fig:si_subset_exploration}). The number of visitors to a location at time \textit{t+1} is similar to the number of visitors at time \textit{t}, revealing stability in visitation patterns and consistency in data collection (Supplementary Fig. \ref{fig:si_pa_evidence}).

\subsection{Land sales}

Decentraland consists of 90,601 unique lands (locations), and each land is minted as a Non-Fungible Token (NFT) on the Ethereum blockchain. It is important to note that the lands on the metaverse were initially sold through an off-chain auction in 2019, and subsequent sales were held on the Ethereum blockchain. We use Etherscan, a block explorer for Ethereum, to extract the data on NFT sales and transfers, finding 15,209 sales of 6,773 land, and 1,562 sales of 1,075 estates (collection of lands) that comprise of 7,159 lands. 

\begin{figure}[ht]
\centering
\includegraphics[width=\linewidth, height = 32em, keepaspectratio]{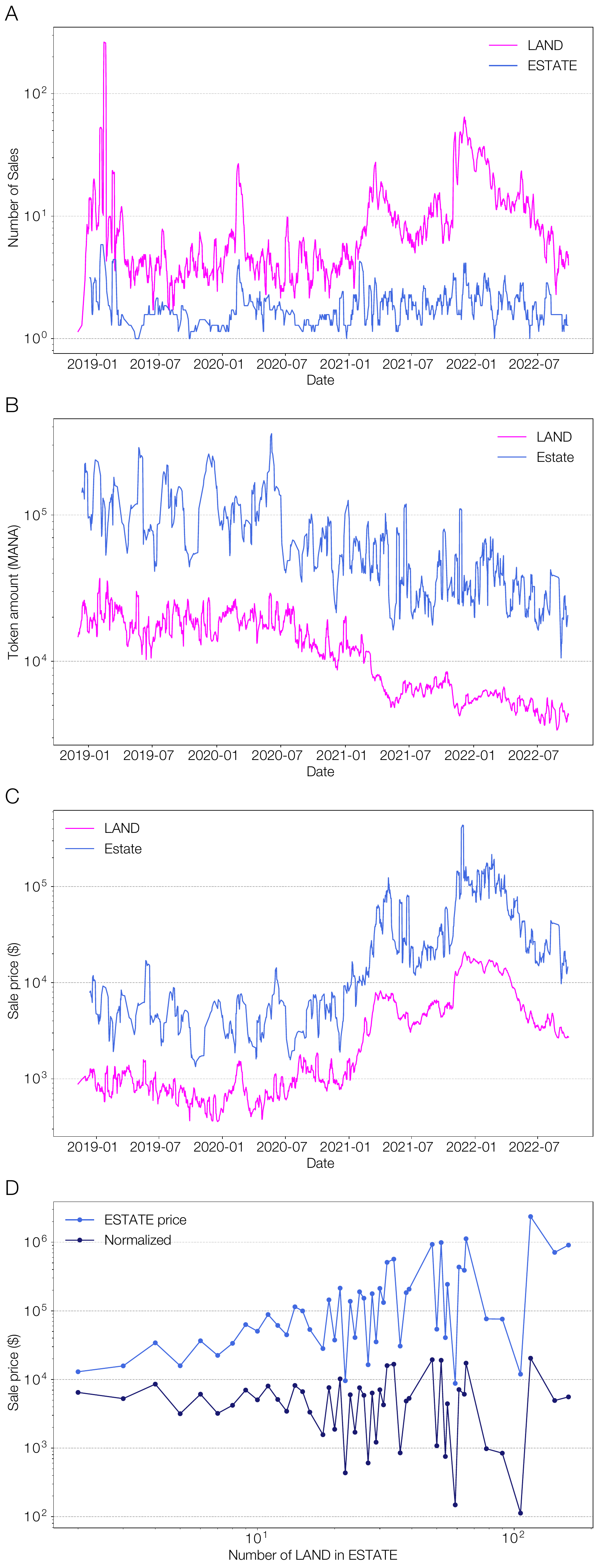}
\caption{\textbf{Pricing trends over time for locations.} We consider the temporal trends in prices for lands and estates (collection of lands sold as a single NFT). \textbf{(A)} Number of location sales over time. We see an uptick in sales in December 2021, but has since stabilized in 2022. \textbf{(B)} Sale price of land based on its MANA price. We find that estates generally sell for considerably higher prices than LAND. \textbf{(C)} Cost of location based on the US Dollar (USD) price. \textbf{(D)} Sale price of estates based on number of lands part of the estate. We find that the value of estates increases based on the number of lands involved. The normalized distribution shows that each land within an estate is valued at a similar rate, irrespective of the size of the estate.}
\label{fig:si_pricing_trends}
\end{figure}

We find that the valuation of land suffered a higher decrease in terms of USD (\$) than in the price of land is terms of MANA (Supplementary Fig. \ref{fig:si_pricing_trends}). In the first week of observation, on average, a LAND sold at 5,290 MANA (\$12,834), while in the last week of observation, an average LAND sold for 4,232 MANA (\$2,982), representing a 20\% decrease in MANA prices and a 76\% decrease in USD valuation. There were 1,884 (12\%) sales of 1,209 (17\%) land parcels that take place during our observation period (Supplementary Fig. \ref{fig:price_visitation_example}). 

\subsection{Contract mobility}

Individuals can also participate in virtual economies, captured by the purchase and selling patterns of NFTs. Each NFT is associated with a specific contract, either of type ERC-721 or ERC-1155 on the Ethereum blockchain. The contracts enforce transaction rules for each NFT sale/transfer, similar to how art galleries set the royalty rates. An individual then uses the contract to purchase a NFT, similar to how collectors purchase a digital art item (NFT) from an art gallery (contract). 

We track the purchasing patterns of the same individuals who also interacted in the virtual world, and collect their transaction history from the two most popular blockchains for NFTs, (a) Ethereum, finding 1,165,310 NFTs from 23,827 contracts collected by 14,732 (9\% of all) users, and (b) Polygon, find 3,112,300 NFTs from 54,918 contracts by 41,870 (25\% of all) users. 

We find that 13,620, 32\% of all who bought NFTs, individuals acquired NFTs from both Ethereum and Polygon. Within this group, 8,926 (65\%) individuals collected more NFTs from Ethereum compared to Polygon. We use the NFT purchases of each individual to build the contract network. For example, the \textit{Foundation} marketplace will be connected to \textit{Artblocks}, if an individual purchased a NFT from \textit{Foundation} and subsequently purchased an NFT from \textit{Artblocks}.

\subsection{Cryptocurrency transactions.} 

\begin{figure}[ht]
\centering
\includegraphics[width=\linewidth, height = 20em, keepaspectratio]{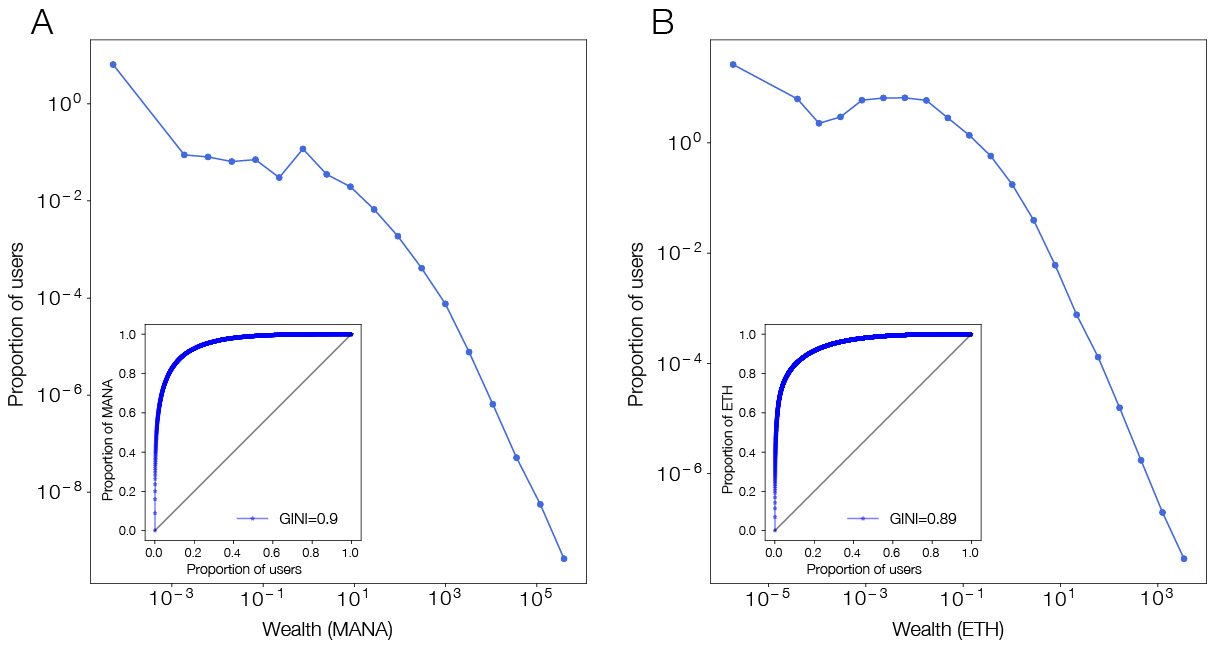}
\caption{\textbf{Wealth distribution of individuals.} We capture the financial status of an individual by analyzing the cryptocurrency within an individual's blockchain wallet. \textbf{(A)} MANA wealth distribution. \textbf{(B)} ETH wealth distribution. The inset panels indicate the cumulative curve of wealth, highlighting that few people own majority of the wealth in the metaverse.}
\label{fig:si_wealth_dist}
\end{figure}

In addition to virtual world mobility and NFT purchases, the blockchain also allows us to track the financial status of the individuals. Each individual account on the metaverse is tied to a blockchain address that is unique and only accessible by the individual. We use the blockchain address of each individual to collect the blockchain history of all users, finding 2,782,023 transactions by 21,553 (13\%) individuals, who collectively own 55,913 ETH, the native currency of Ethereum (Supplementary Fig. \ref{fig:si_wealth_dist}). We also extract transactions involving MANA, the native currency of Decentraland that exists as an ERC-20 token on the Ethereum blockchain, finding 1,056,914 transactions by 5,596 (3.5\% of all) individuals representing 2,240,012 MANA. Overall, 3,887 (2.3\% of all) users own both MANA and ETH, while 17,685 (10\%) only own ETH and 1,709 (1\%) only own MANA. 

The data to collect latest data is provided on Github at https://github.com/Barabasi-Lab/metaverse-mobility

\section{Spatial visitation and selling price}

In the physical world, visitation patterns creates urban centers, representing geographical clusters that attract high number of visitors. As the metaverse lacks the inherent physical features and the commuting costs associated with movement, we ask, do the virtual world also result in spatial clusters of visitors? and do the visitation patterns also drive the economic value of the location? 

\begin{figure}[ht]
\centering
\includegraphics[width=\linewidth, height = 20em, keepaspectratio]{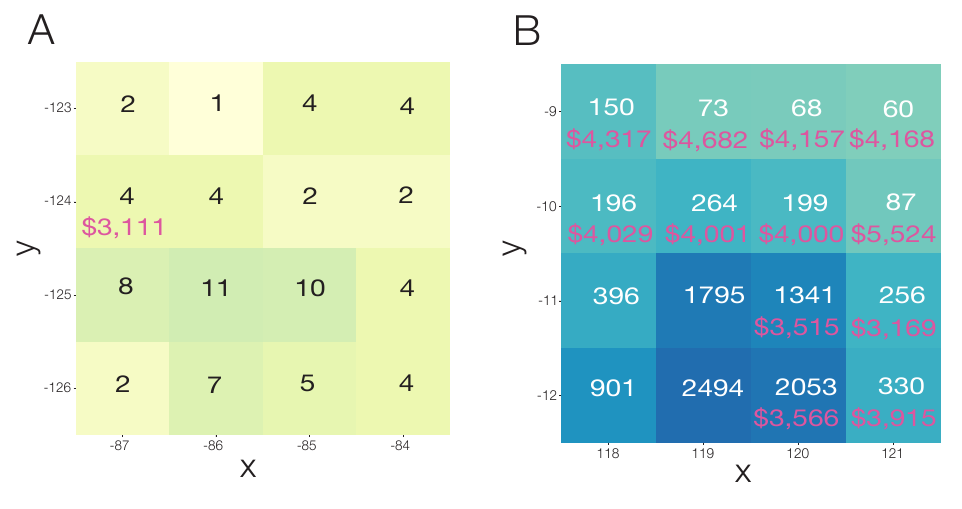}
\caption{\textbf{Example of lands along with their visitation and sale prices.} We highlight the exploration patterns in two different regions in the virtual world, one with \textbf{(A)} low visitation and another with \textbf{(B)} high visitation. The panels indicate that visitation numbers are not correlated with sale price of lands.} 
\label{fig:price_visitation_example}
\end{figure}

\begin{figure}[ht]
\centering
\includegraphics[width=\linewidth, height = 20em, keepaspectratio]{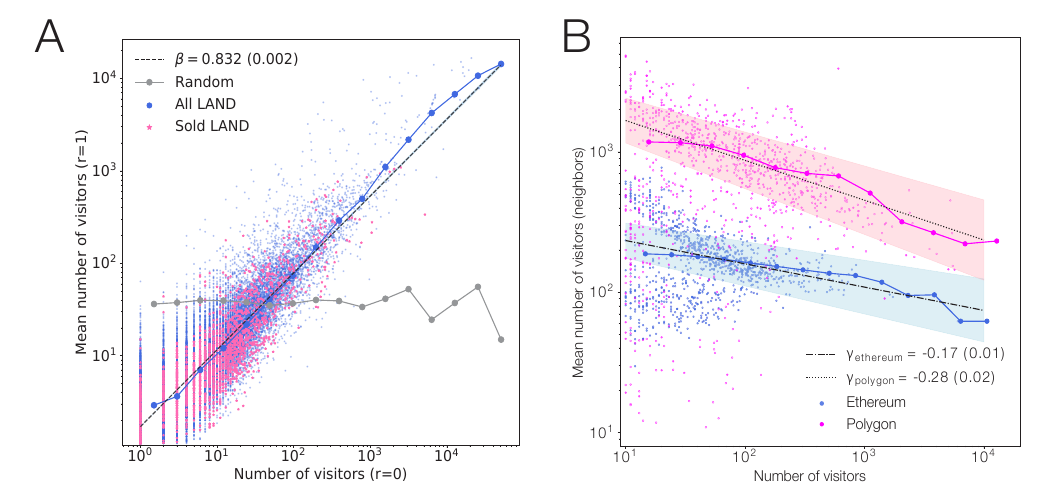}
\caption{\textbf{Spatial segregation in visitation.} \textbf{(A)} Mean number of visitors in the neighborhood of a land. We find that the number of visitors a land receives is highly correlated, on average, to the number of visitors its neighboring land receives ($\beta = 0.82$, $R^{2} = 0.78$). Such clustering of visitation based on spatial distance is not visible if visitors were randomly distributed across all lands. We conduct randomization by generating synthethic visitation trajectories of all individuals while keeping the number of visited locations constant. \textbf{(B)} Mean number of visitors in a contract compared to the average number of visitors in their network neighborhood. We find that the number of visitors in a contract is negatively correlated with the number of visitors in its neighborhood. }
\label{fig:neigh_visitors_si}
\end{figure}

To understand the spatial concentration of visitors in the virtual world, we measure the number of visitors received by a particular land and its surrounding areas. We find that the number of visitors a land ($N_S$) receives scales with the number of visitors its neighboring land receives ($N_{SS}$), following $N_{S} \propto \langle N_{SS}\rangle^{\beta}$, where $\beta=0.83$ (CI: 0.828, 0.835, $R^{2} = 0.78$, Supplementary Fig. \ref{fig:neigh_visitors_si} A). For example, a popular land (-99,128) received 15,251 visitors and its neighboring parcels, on average, received 13,757 visitors. Similarly, a less popular land (-86,-124) receives 4 visitors and its neighboring parcels, on average, receives 5 visitors (Supplementary Fig. \ref{fig:price_visitation_example}). In other words, the spatial dynamics of mobility in the virtual creates clusters of highly visited virtual locations. 

\begin{figure}[ht]
\centering
\includegraphics[width=\linewidth, height = 20em, keepaspectratio]{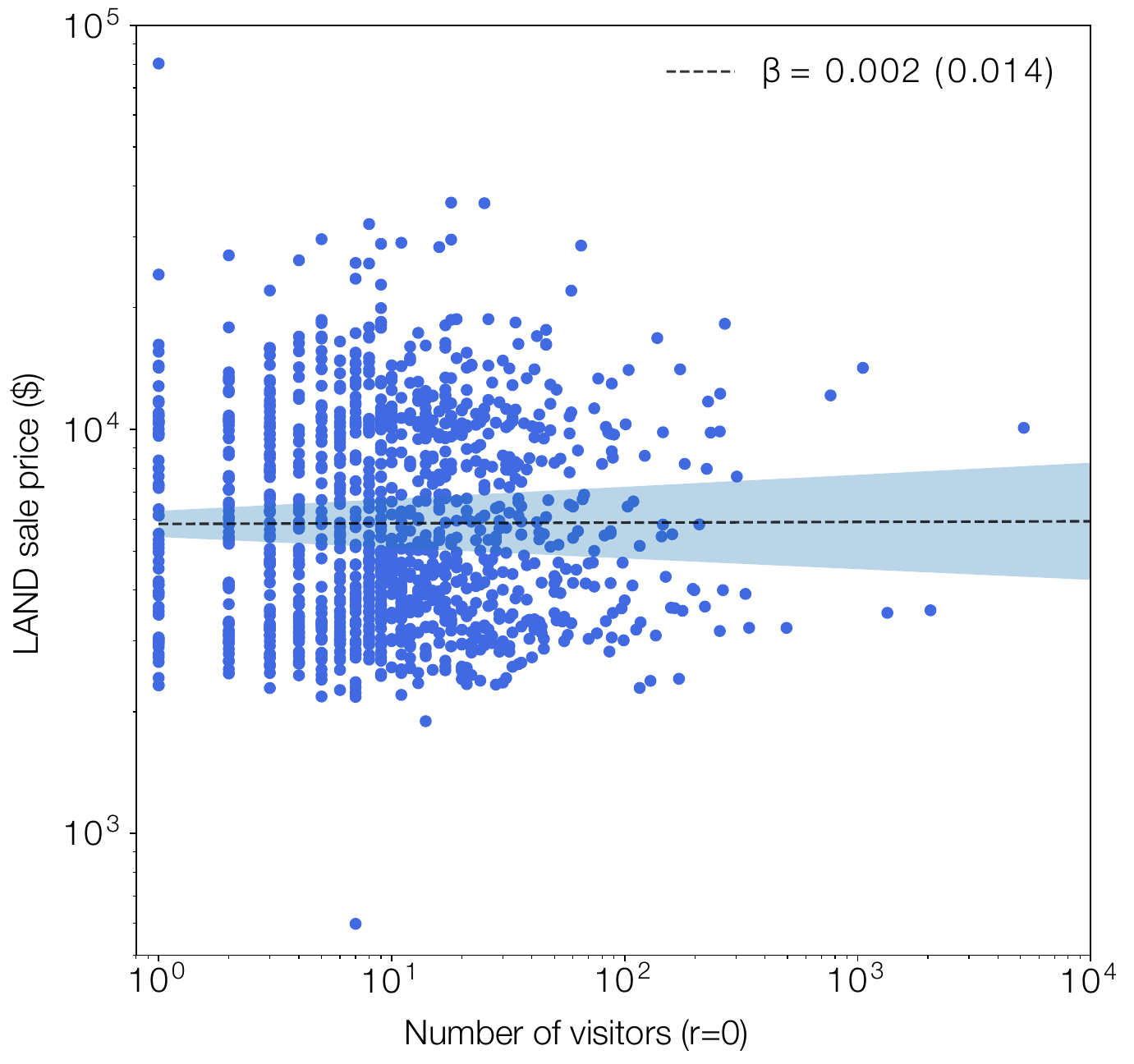}
\caption{\textbf{Selling price of land and visitation patterns}. We find that the number of visitors a land receives is not correlated with the sale price of the land ($\beta = 0.002$, $R^{2} = 0$). This indicates that visitation patterns does not affect selling patterns. Parenthesis indicate the standard error of the estimates. }
\label{fig:visitor_sale_price_land_si}
\end{figure}

\begin{figure}[ht]
\centering
\includegraphics[width=\linewidth, height = 20em, keepaspectratio]{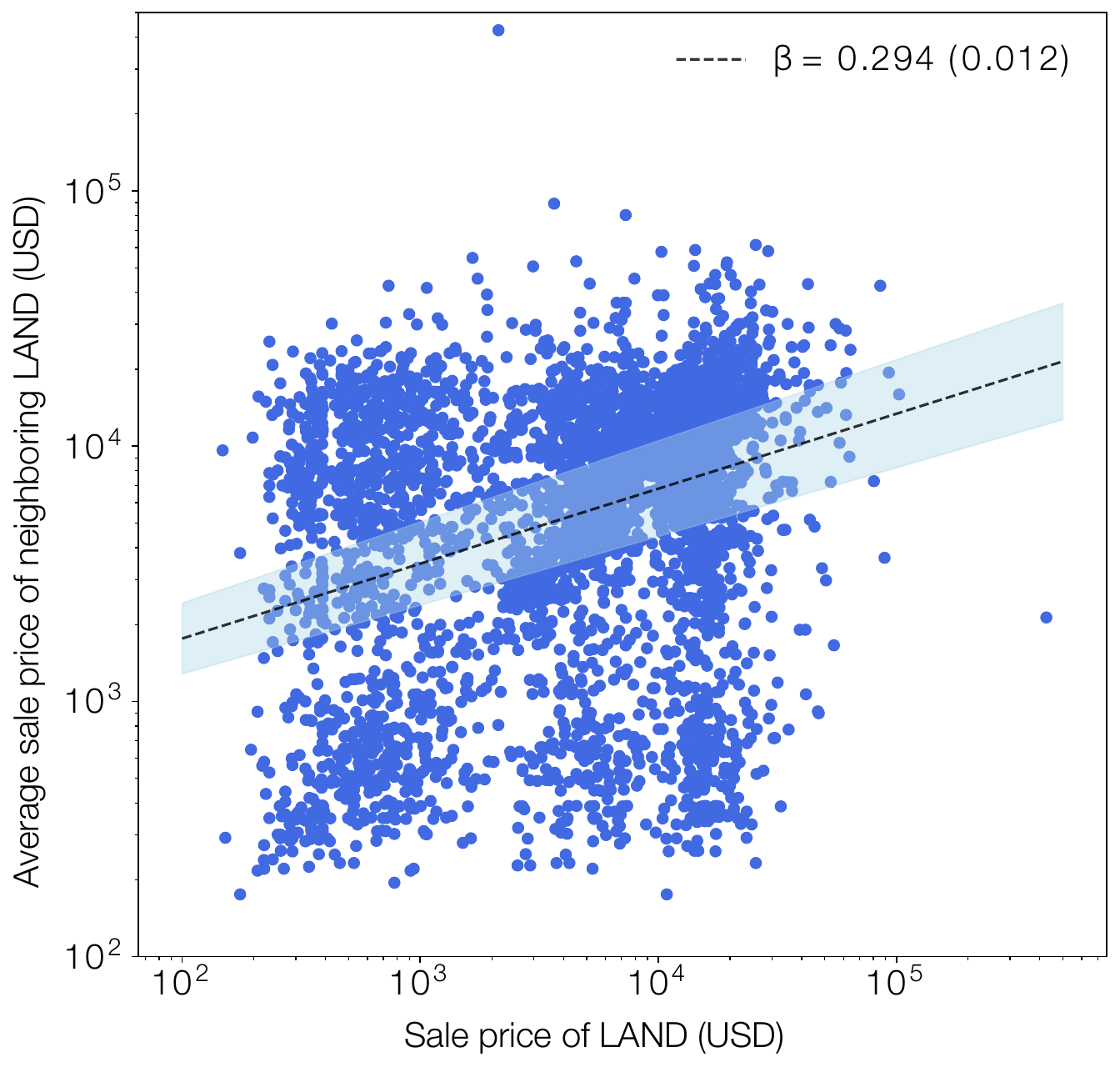}
\caption{\textbf{Spatial segregation in sale price.} We consider the sale price of a land and find the sale price of its neighboring lands. We find a weak correlation between the selling price of a location and the demand price in its neighborhood ($\beta = 0.294$, $R^{2} = 0.1$). This suggests that, unlike the physical world where expensive neighborhoods cluster together, the metaverse lacks a spatial concentration of highly expensive neighborhoods. }
\label{fig:neigh_sale_price_land}
\end{figure}

Next, we measure the number of visitors a land receives and the selling price of the location. We find that the visitation numbers of a location is not associated with the selling price of the land (Supplementary Fig. \ref{fig:visitor_sale_price_land_si}, $R^{2} = 0$). We also fail to discover a strong relationship between the price of a location and the average price of its neighboring locations (Supplementary Fig. \ref{fig:neigh_sale_price_land}, $R^2 = 0.1$). This represents a contrast to the physical world, where the pricing of a location is influenced by foot traffic to that location and its neighboring areas\cite{west2018scale_si}. 

Overall, we find that the mobility patterns of individuals creates a regions of high visitation concentration in the metaverse, and the pricing dynamics of the locations are not influenced by visitation numbers. 

\section{Individual mobility patterns}

\subsection{Number of visited locations}

\begin{figure}[ht]
\centering
\includegraphics[width=\linewidth, height = 20em, keepaspectratio]{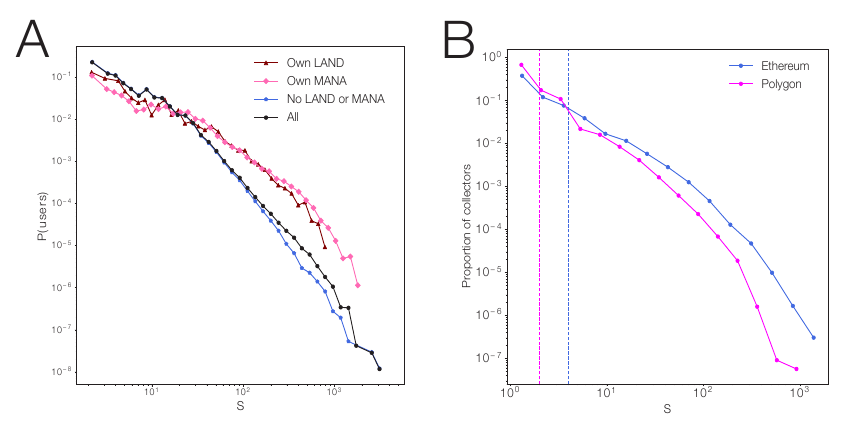}
\caption{\textbf{Number of visited locations.} \textbf{(A)} Virtual world. We find that users that do not own any cryptocurrency or land are more exploratory than users with financial investment in the metaverse. \textbf{(B)} Contract mobility. We show the number of contracts that an individual has interacted with when purchasing an NFT. The dashed line indicates the median number of contracts.}
\label{fig:s_explored_si}
\end{figure}

To understand individual mobility patterns, we first begin by quantifying the number of locations visited by each individual. In virtual world, visitation is represented as the total number of lands visited, and in the contract space this is quantified by the number of contracts from which they purchased one or more NFTs. We find that the number of locations visited by each individual in the metaverse displays significant variations, captured as a fat-tailed distribution in the number of locations visited $P(S)$ (Supplementary Fig. \ref{fig:s_explored_si}).

Specifically, an average individual only visited 18 lands within the virtual world, representing less than 1\% of all lands (Supplementary Fig. \ref{fig:s_explored_si} A). Similarly, an average individual bought NFTs from 20 different contracts on Ethereum, and an average individual on Polygon purchased from 8 contracts (Supplementary Fig. \ref{fig:s_explored_si} B). These patterns suggest that despite the lack of physical and time restrictions to discover and explore new locations, individuals tend to focus their mobility to a small fraction of the metaverse. 

\subsection{Time allocation in different locations}

\begin{figure}[!ht]
\centering
\includegraphics[width=\linewidth, height = 20em, keepaspectratio]{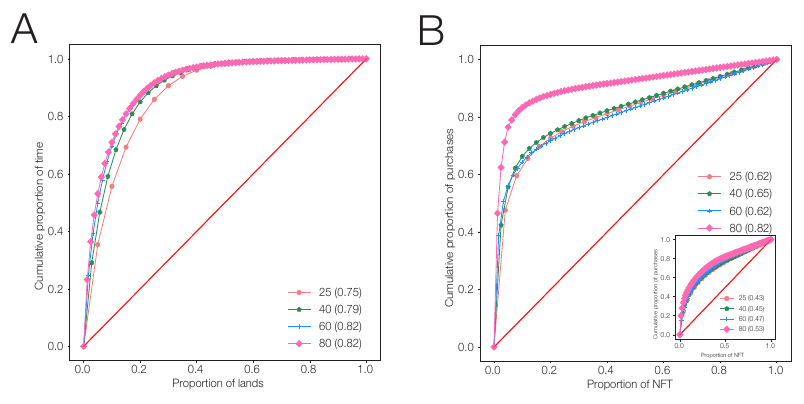}
\caption{\textbf{Inequality in land and contract visitation.} We use the GINI coefficient to measure inequality in visitation patterns across all visited locations. A score of 1 indicates complete inequality and a score of 0 indicates equal visitation numbers across all visited locations. \textbf{(A)} The visitation curve for virtual world. \textbf{(B)} The visitation curve for the Polygon blockchain, inset shows the same curve for the Ethereum blockchain. We find that individuals tend to disproportionately allocate time and resources across all visited locations. Parenthesis indicates the GINI coefficient.}
\label{fig:si_gini}
\end{figure}

We examine the distribution of time spent at all visited locations. We find that irrespective of number of visited locations, individuals spend almost 80\% of their time at 20\% of the locations in the virtual world, characterized by a high GINI coefficient (0.75-0.82) (Supplementary Fig. \ref{fig:si_gini} A). Similarly, individuals display purchase most of their NFTs from a few contracts (GINI: 0.62-0.82), a pattern that is consistent across the Ethereum and Polygon blockchains (Supplementary Fig. \ref{fig:si_gini} B). 

\begin{figure}[!ht]
\centering
\includegraphics[width=\linewidth, height = 20em, keepaspectratio]{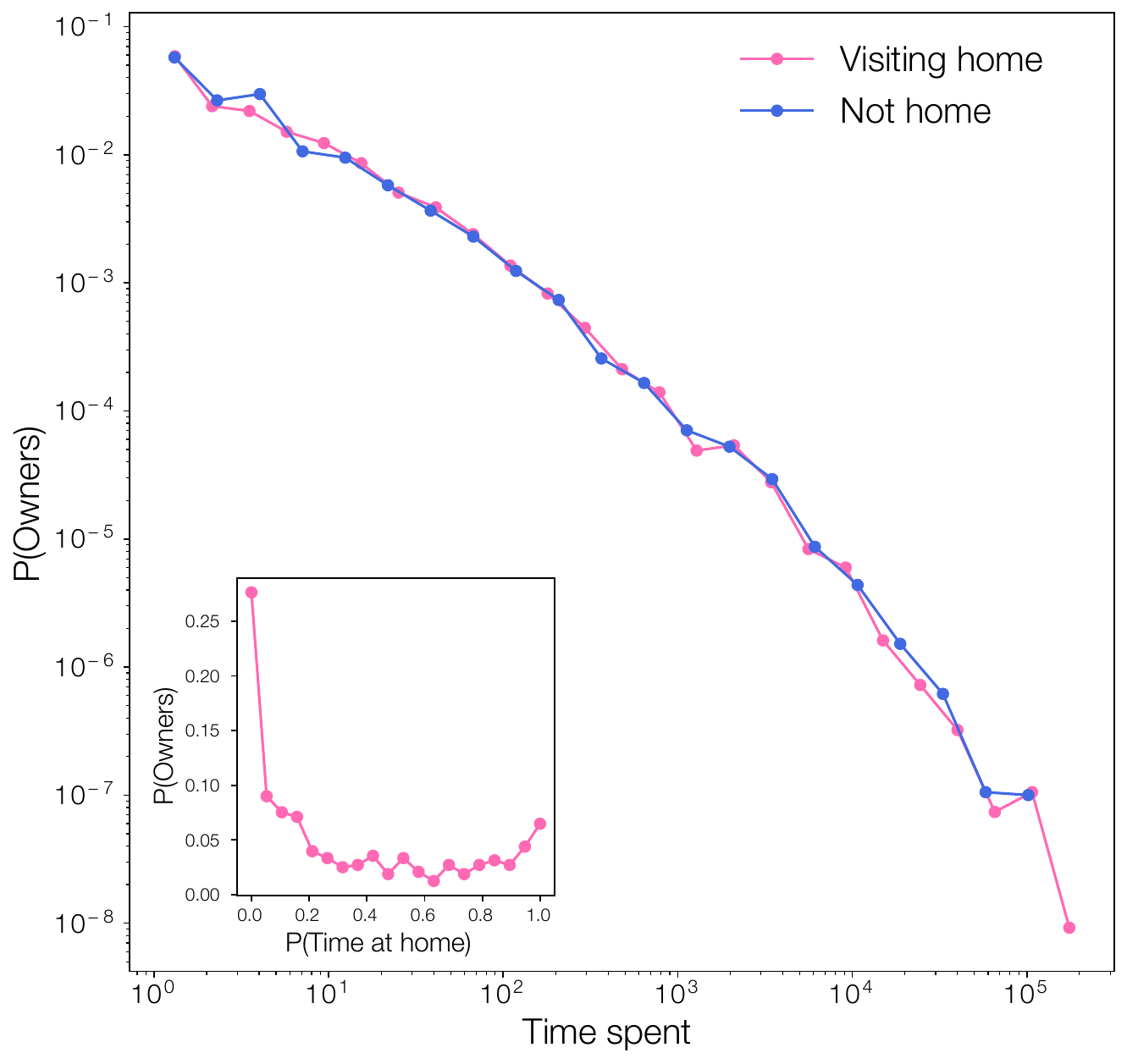}
\caption{\textbf{Time spent at the owned land.} We consider all of the individuals that own a land and measure the time spent at that location. We find that over 65\% of the land owners spend more than half of their time at the owned land.}
\label{fig:si_time_spent_home}
\end{figure}

In the physical world, visitation patterns are influenced by the location of their work and home. To test the role of home in the virtual world, we examine the mobility patterns of land owners, asking how often land owners visit their owned location? We find that more than 65\% of the land owners spend more than half of their time at the owned location (Supplementary Fig. \ref{fig:si_time_spent_home}). This suggests that the presence of an owned land in the virtual world influences the hangout location in the virtual world. 

\begin{figure}[!ht]
\centering
\includegraphics[width=\linewidth, height = 20em, keepaspectratio]{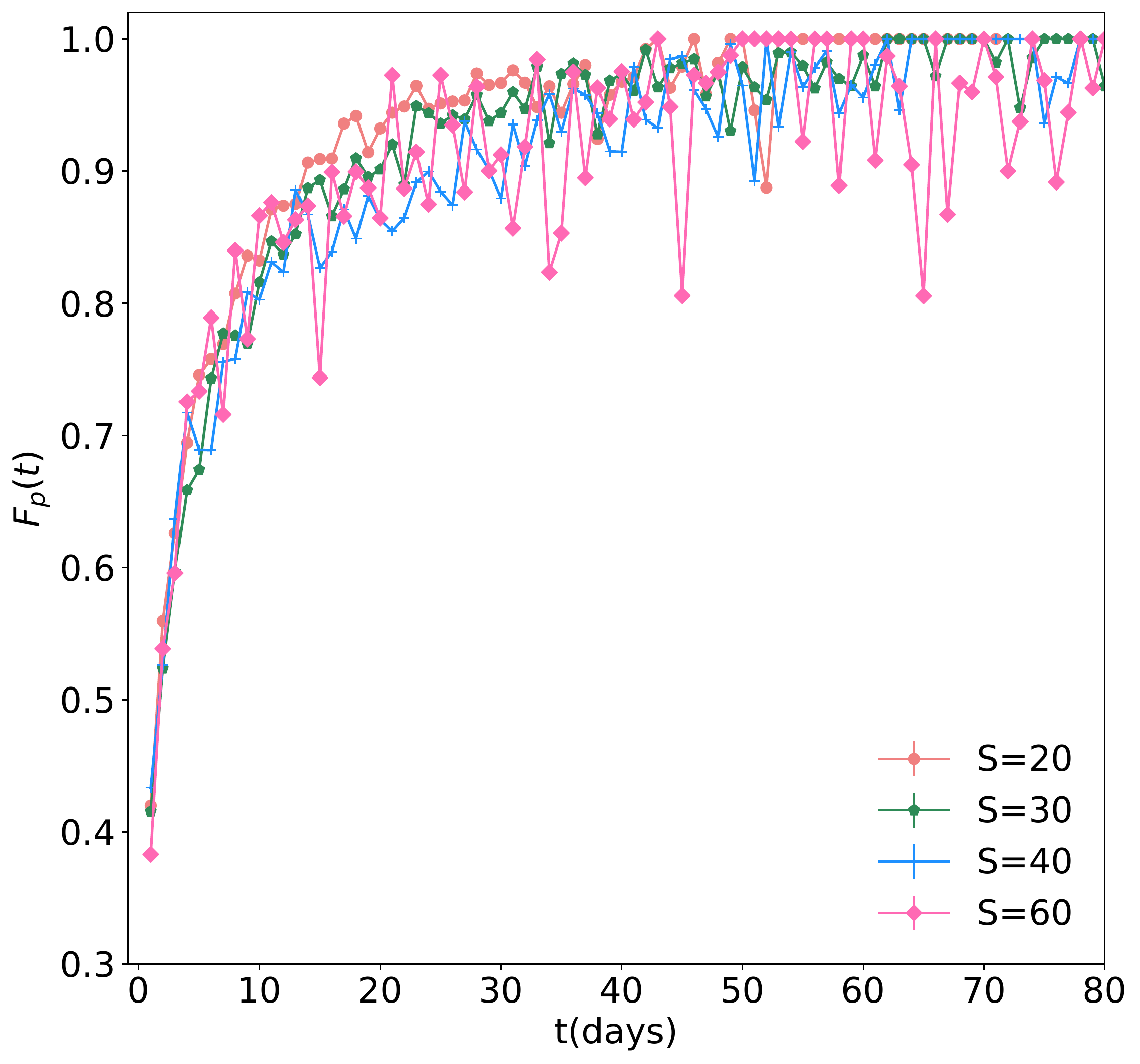}
\caption{\textbf{Location return probability distribution.} We examine the probability that an individual returns to a previously visited locations on each subsequent day. We find that individuals tend to return to their previously visited locations with high probability, irrespective of the number of visited locations ($S$). This suggests that individuals develop a preference towards a few locations.}
\label{fig:fpt_individual}
\end{figure}

As only a small fraction of users own a land, we ask, do non-land owners display such affinity to specific locations? We examine the probability that an individual returns to a previously visited locations on each subsequent day. We find that individuals tend to return to their previously visited locations with high probability (over 90\%), irrespective of the number of lands visited (Supplementary Fig. \ref{fig:fpt_individual}). 

Overall, the analysis shows that individuals display an attraction towards a few locations, prompting them to spend uneven fraction of their time and frequently return to their preferred location. 

\subsection{Capturing jump distances}

\textbf{Virtual world.} All of the locations in the Decentraland metaverse are placed on a matrix comprising of 301 rows and 301 columns. To measure the distance between two lands, we use the Manhattan distance metric, calculated as the sum of absolute difference between two locations in the x and y coordinates. For example, if parcel 1 is located at point ($x_1$, $y_1$), and parcel 2 is located at point ($x_2$, $y_2$), the distance between the two points will be: 

\begin{equation}
    d(1, 2) = |x_1 - x_2| + |y_1-y_2|,
\end{equation}

\begin{figure}[ht]
\centering
\includegraphics[width=\linewidth, height = 20em, keepaspectratio]{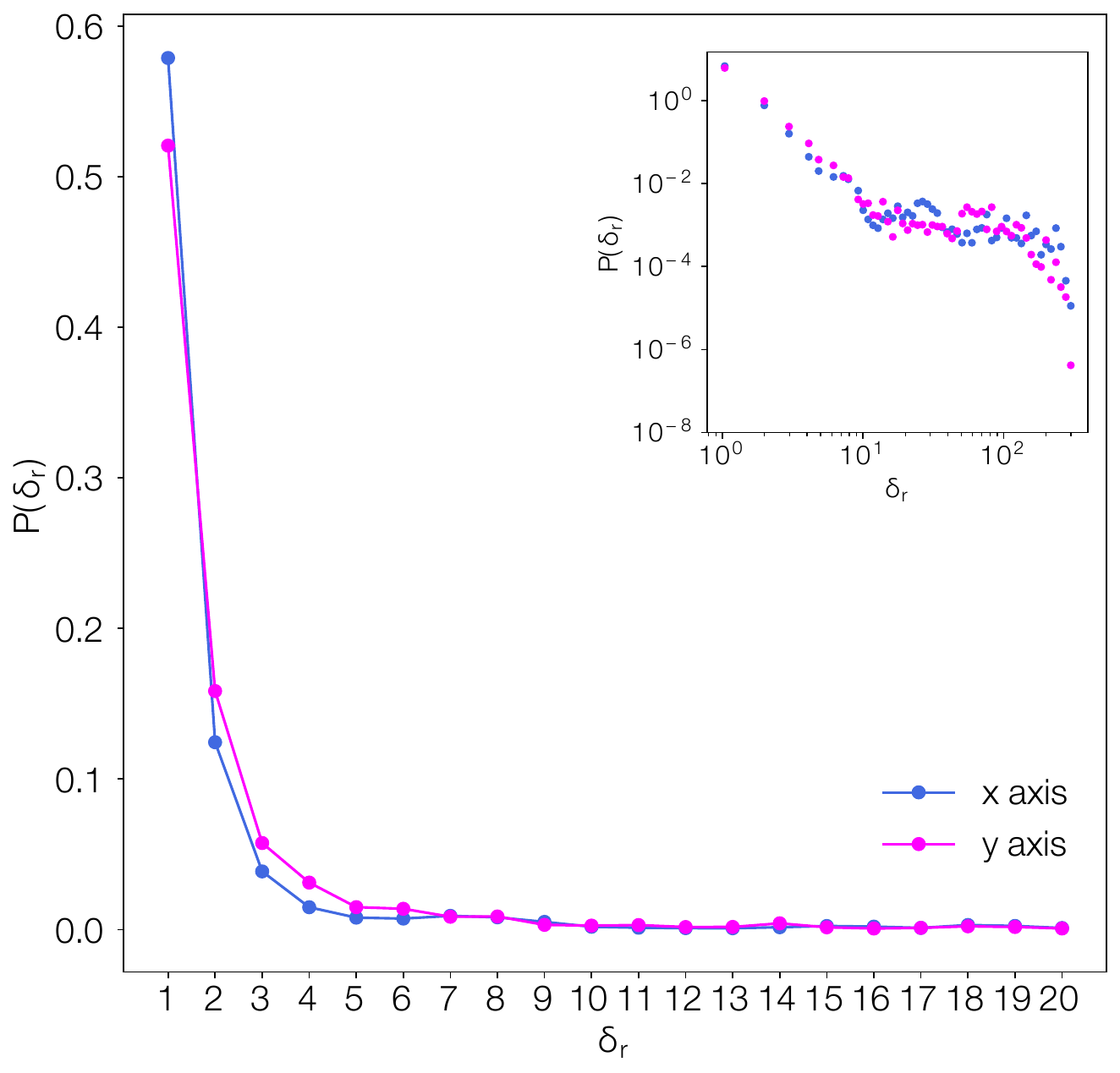}
\caption{\textbf{Jump distance based on coordinates.} The virtual world exists on a $x-y$ plane suggesting that the movement could occur in either left/ right direction ($x$ axis movement) or up/ down direction ($y$ axis movement). We show the jump distribution along each of the axis, where a $\delta_r = 2$ in the $x$ axis indicates a left/ right movement along the $x$ axis. The plot shows that the jump distance metric is not influenced by movement towards a specific direction.}
\label{fig:si_xy_mobility}
\end{figure}

\begin{figure}[ht]
\centering
\includegraphics[width=\linewidth, height = 20em, keepaspectratio]{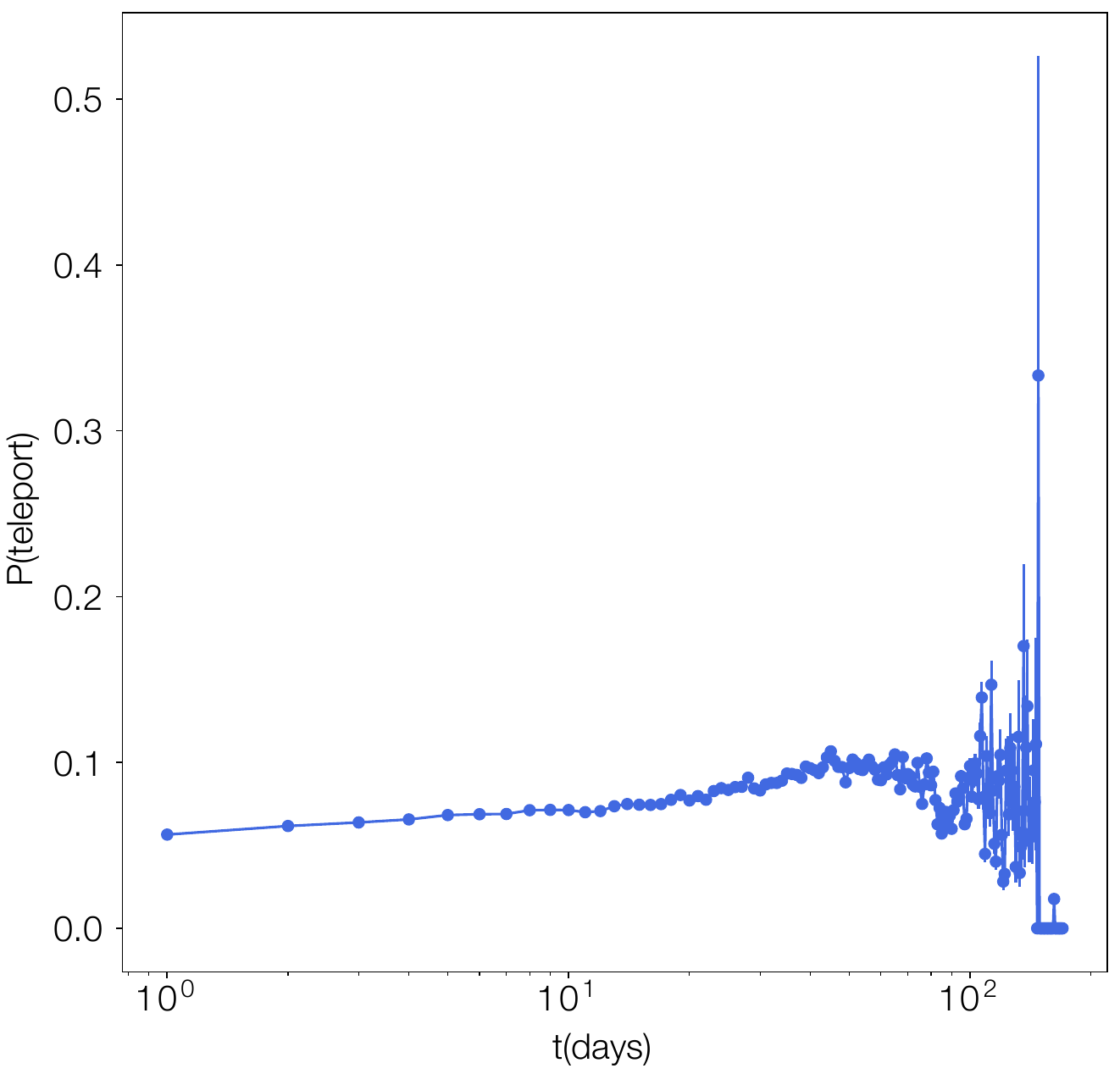}
\caption{\textbf{Teleportation in the metaverse.} We consider teleportation to indicate any distance displacement greater than $\delta_r > 10$, where the individual would prefer to teleport rather than walk along the path. We find that individuals teleport 10\% of the visits each day in the metaverse. Bars indicate standard error.}
\label{fig:pr_teleport_days}
\end{figure}

We find that individuals rarely tend to make long distance jumps (Supplementary Fig. \ref{fig:si_xy_mobility}). In fact, only 10\% of the movements accounts for a distance greater than 10 (Supplementary Fig. \ref{fig:pr_teleport_days}), that remains consistent during each new visitation day. Despite the freedom to engage in large displacements and teleportation, we find that individuals rarely engage in large displacements. 

\begin{figure}[ht]
\centering
\includegraphics[width=\linewidth, height = 20em, keepaspectratio]{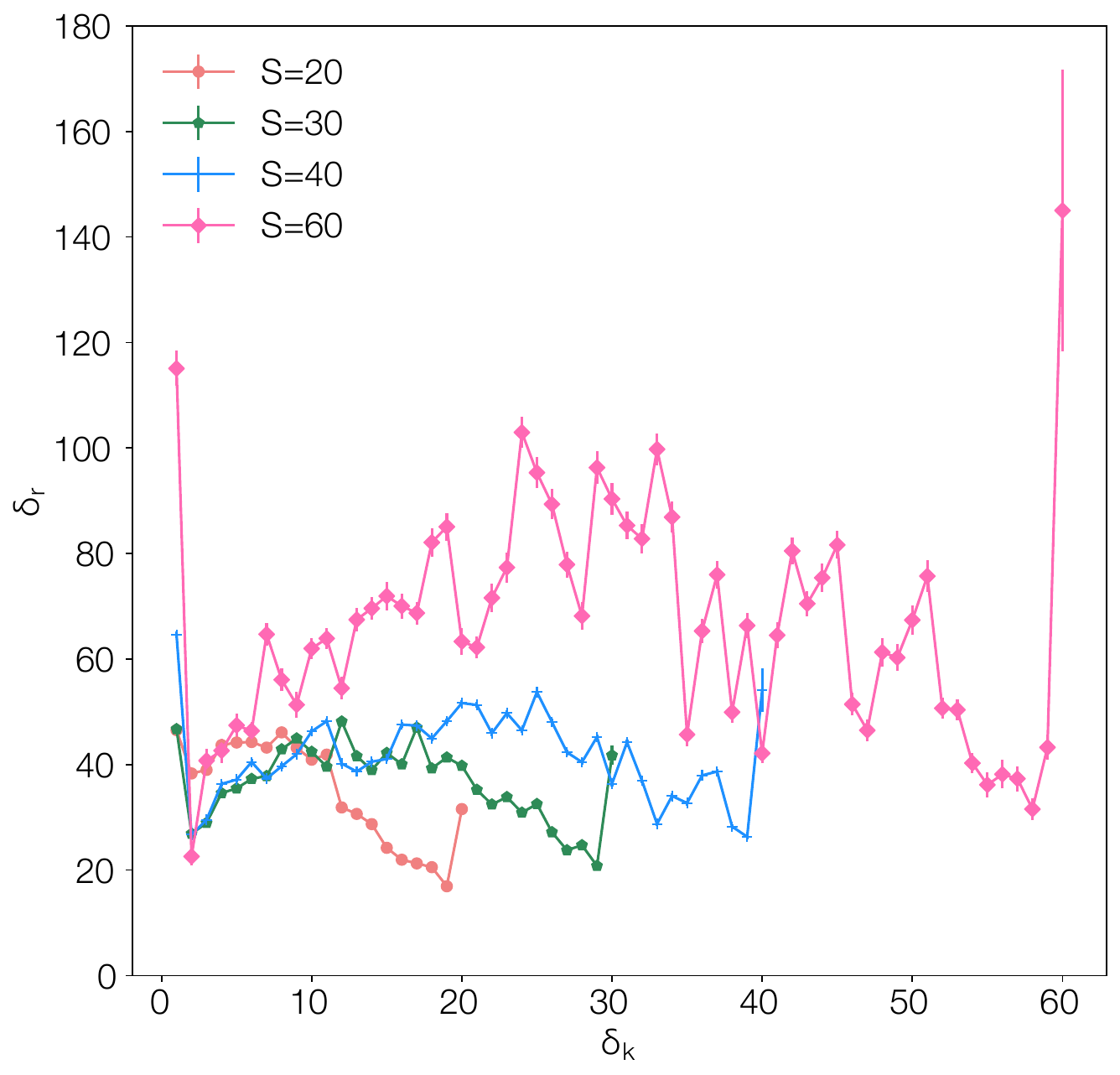}
\caption{\textbf{The distance between the highly visited locations.} We rank the visited locations for each individual based on the number of visits to that location ($\delta_k$), and study the distance between the ranked visited location ($\delta_r$). A distnace of $\delta_r = 10$ for $\delta_k=2$ suggests that the second most visited locations is at a distance of 10 from the first most visited location. We find that individuals in the metaverse tend to identify favorite locations that are located at farther distances from each other, irrespective of number of visited locations ($S$). This acts as a distinguishing feature of the metaverse where trip frequency is not constrained by commuting time and costs.}
\label{fig:delta_r_delta_k}
\end{figure}

In the physical world, the spatial component of mobility (physical distances) affects the individual visitation patterns. It has been shown that the least visited locations linearly decreases with distance from home\cite{schlapfer2021universal}, suggesting that trip frequency is strongly influenced by physical distance, often determined by commuting costs and time. Curious to investigate whether a similar feature emerges in the metaverse, we compare the distances between the ranked list of highly visited locations. We find that individuals in the metaverse often frequently visit locations that are located at farther distances and do not exhibit a characteristic decay in likelihood to trip frequency based on distance (Supplementary Fig. \ref{fig:delta_r_delta_k}).

\textbf{Contract mobility.} To capture the jump distance in the contract space we utilize the contract network (CN), where the nodes are contracts and a link signifies movement between the contracts. We measure distance between contracts using the network distance\cite{newman2018networks}, captured via the shortest path length between the two explored locations, $\delta_r = d(a,b)$. For example, the distance between the two contracts, $d(a,b)$, is one if they are directly connected to each other. The distance would be 2 if the two contracts share a common neighbor. 

\section{Building the network}

\textbf{Virtual World Network (VWN).} The virtual world network is created by following the spatial trajectory of individuals in the virtual world. We consider a link between two locations A and B, if an individual visited location A and then subsequently visited location B. The time-resolved version of the network allows us to estimate the movement patterns across spatial locations in the form of a network. The mobility network in the virtual world contains 69,433 nodes and 618,358 links.

\begin{figure}[ht]
\centering
\includegraphics[width=\linewidth, height = 20em, keepaspectratio]{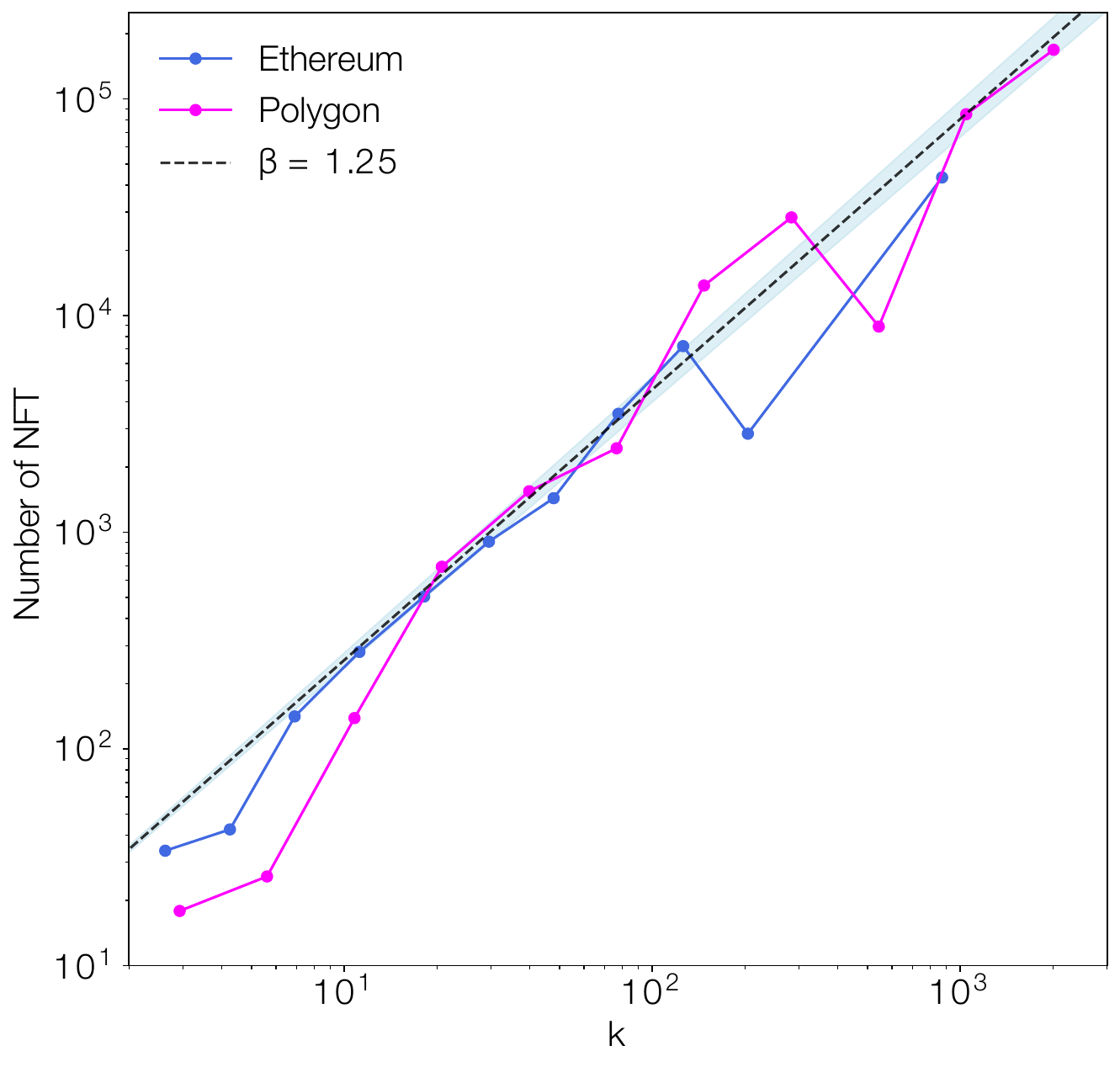}
\caption{\textbf{Network effects in number of NFT sold.} We show the number of NFTs sold by a specific contract (location) and its degree in the contract mobility network (CN). The fact that $\beta>1$, suggests a super linear relationship between the centrality of the location and the number of sales. }
\label{fig:si_deg_nft_sold}
\end{figure}

\textbf{Contract network (CN).} The contract network is created by following the NFT purchases of individuals across contracts. Any two contracts A and B would have a link if individual purchased an NFT from contract A and then subsequently purchased another NFT from contract B. This time-resolved approach allows us to understand the temporal flow between contracts. We construct two networks, one for Ethereum finding 13,787 nodes and 198,126 links, and another for Polygon 16,897 nodes and 96,536 links. We find that the degree of a contract in the network is positively correlated with the number of NFTs the contract sells ($\beta=1.25$, Supplementary Fig. \ref{fig:si_deg_nft_sold}). The fact that $\beta>1$, indicates a super linear scaling between the location's centrality and the number of sales, suggesting that the network structure captures node importance and its ability to sell NFTs. 

\subsection{Randomizing the network}

\begin{figure}[ht]
\centering
\includegraphics[width=\linewidth, height = 20em, keepaspectratio]{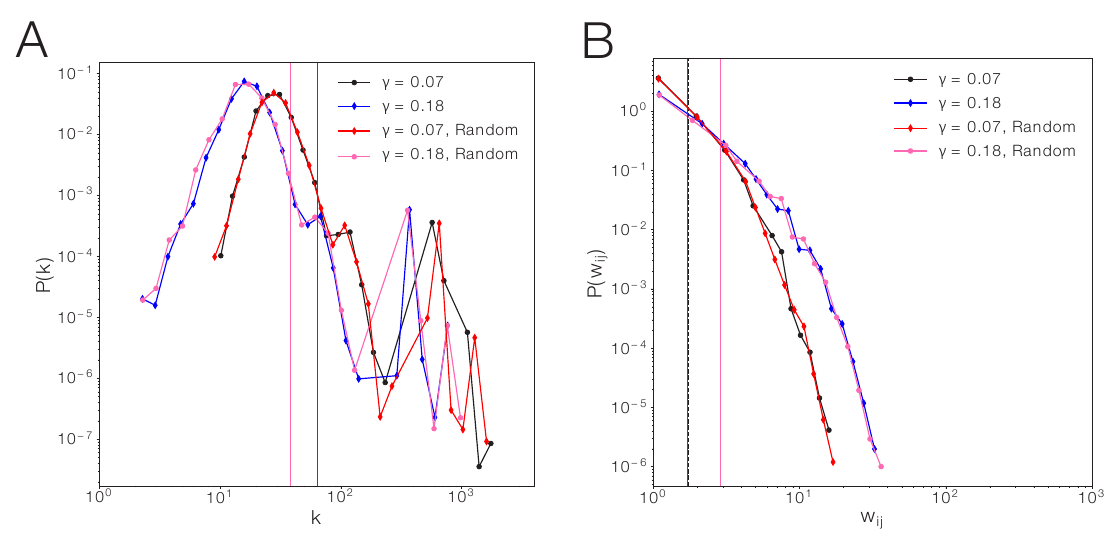}
\caption{\textbf{EPR model and random simulations.} We consider the model parameters from empirical data and simulate the individual trajectories. We consider two models: (a) the EPR model and (b) the random model, where individuals randomly choose locations. \textbf{(A)} The degree distribution $P(k)$. We find that both the EPR model and the Random model are unable to uncover the empirically observed fat-tailed degree distribution. \textbf{(B)} The link weight distribution $P(w_{ij})$. We find that the EPR model is unable to uncover the heterogeneous flows between locations. Note that the empirical data contains link weights of the order $w_{ij} \sim 10^3$.}
\label{fig:si_epr_net_dist}
\end{figure}

To characterize the mechanisms behind the empirical observations, we construct a random network using synthetic user trajectories. We keep the total number of visited locations constant for each individual and conduct location randomization, where we randomly sample locations from all possible locations. The randomization follows the procedure where individuals are neither biased towards locations either based on their own visitation history nor through popularity of the locations, allowing us to break the temporal and user-location correlations. We follow this procedure for all individuals and generate synthetic trajectories. The results show that the random model fails to discover heterogeneous flows between the locations (Supplementary Fig. \ref{fig:si_epr_net_dist}). 


\section{Modeling Mobility}

\subsection{Empirical mechanisms}

\begin{figure}[ht]
\centering
\includegraphics[width=\linewidth, height = 20em, keepaspectratio]{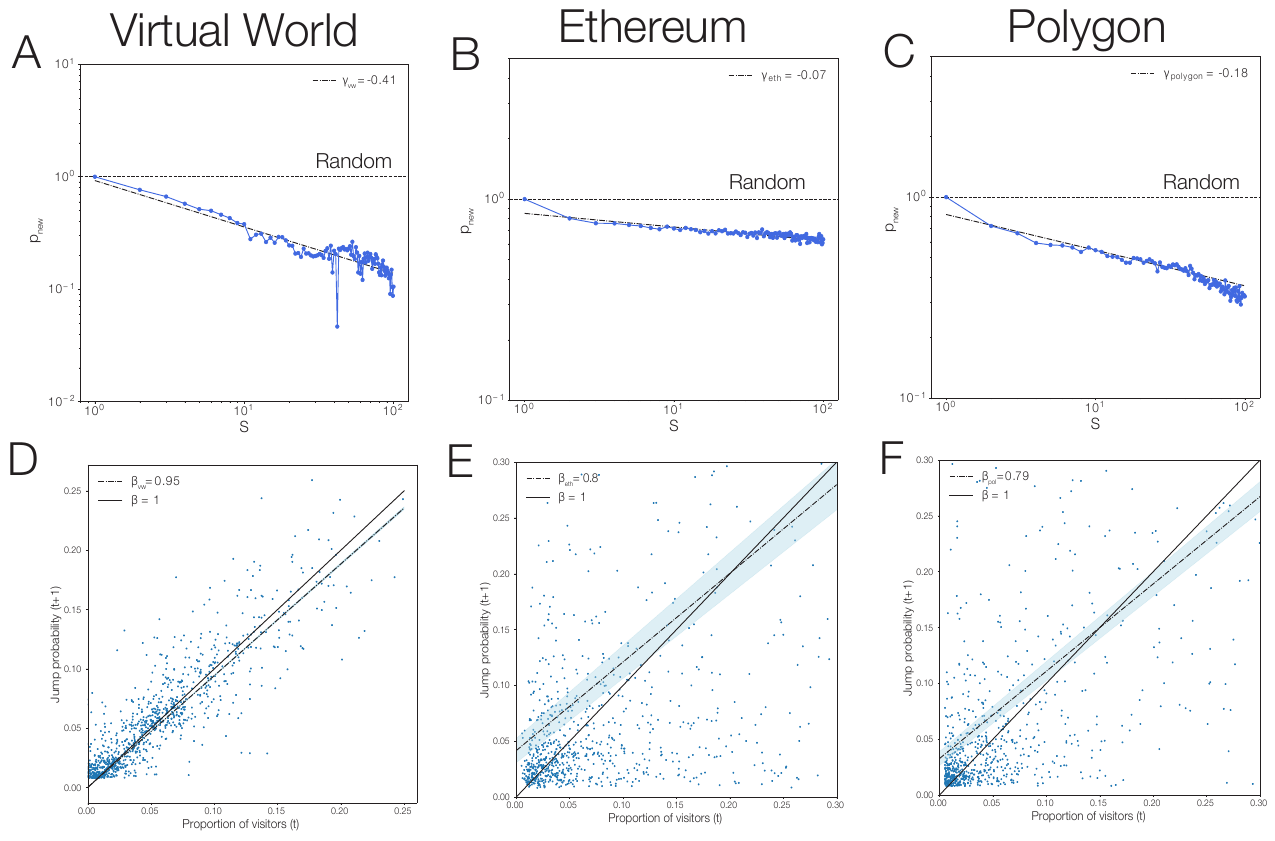}
\caption{\textbf{Mechanisms of individual mobility.} \textbf{(A-C)} We measure the probability of visiting a new location, $p_{new}$, as a function of number of previously visited locations $S$. We find that the distribution is well-approximated as $p_{new} \propto S^{-\gamma}$, where $\gamma_{vw} = 0.41$, $\gamma_{ethereum} = 0.07$, and $\gamma_{polygon} = 0.18$. \textbf{(D-F)} We measure the probability that an individual will move to a specific location as a function of the normalized number of visitors at that location. We observe a linear relationship between the two variables confirming the preferential movement hypothesis. }
\label{fig:model_specification_si}
\end{figure}

The components of the model is inspired by two fundamental features about individual mobility: 

(a) We compare the probability of visiting a new location based on past visitation history. We find that the probability that an individual visits a new location decreases with number of past visits following $p_{new} \propto S^{-\gamma}$ (Supplementary Fig. \ref{fig:model_specification_si} A-C) with $\gamma_{vw} = 0.41$ in the virtual world and $\gamma_{ethereum} = 0.07$ and $\gamma_{polygon} = 0.18$, This shows that an individual's inclination to visit new locations decrease over time. 

(b) We test the effect of popularity on the individual visitation probabilit by comparing the proportion of visitors at a given time step \textit{t} and measure the likelihood that an individual will move to that location at the next time step \textit{t+1}. The empirical data indicates a linear relationship between the two variables, suggesting that an individual is more likely to move to a popular location (Supplementary Fig. \ref{fig:model_specification_si} D-F). 

As these two findings are applicable in both virtual world and the contract space, it serves as the fundamental drivers of human mobility. 

\subsection{EPR Model}

To understand how well the existing models can account for mobility in the metaverse, we start from the well-established Exploration and Preferential Return (\textit{EPR}) model\cite{song2010modelling}, a model that has emerged as a foundational framework for human mobility. In this model, given the prior visiting trajectory of $S$ locations for each individual, at each time step, an individual has two choices: with probability $p_{new} = \rho S^{-\gamma}$, it decides to explore a new random location drawn from the distance function, $P(\delta_r)$. Otherwise, with probability $1-p_{new}$, it decides to revisit a previously visited location, and the location is selected proportional to the number of past visits, $\pi_i(i) = \frac{m_{i}}{\sum_j m_j}$. 

\subsection{m-EPR Model}

In the \textit{m-EPR} model, given the visiting trajectory of $S$ locations, at each time step, an individual has two choices: with probability $p_{new} = \rho S^{-\gamma}$, it explores a new location sampled based on the aggregate number of visits to that location by it, $\pi(i) = \frac{m_i}{\sum_j m_j}$. With probability $1-p_{new}$, it revisits a previously visited location, again proportional to the aggregate number of visits to the location by all users, $\pi(i) = \frac{m_i}{\sum_j m_j}$. The model has two key differences from the \textit{EPR} model: 

(a) The individual movements are not limited based on the distance between the locations. That is, the \textit{m-EPR} model is a distance-agnostic explanation of human mobility. 

(b) The individual movements are affected by the popularity of each location rather than the number of individual visits. That is, the \textit{m-EPR} model introduces a novel popularity driven mechanism for individual decisions, not detected before in physical mobility. 

\subsection{Analytical predictions}

\subsubsection{Number of visited locations ($S$)}

In the model formulation, $p_{new} = \rho S^{-\gamma}$, where $S$ is the number of previously visited locations. This allows us to formulate a differential equation of the form: 

\begin{equation}
    \frac{dS}{dn} = \rho S^{-\gamma}.
\end{equation}

By integrating the equation, we obtain

\begin{equation}
\begin{split}
S^{1+ \gamma} = \rho n(1+\gamma)\\
S = (\rho n)^{\frac{1}{1+\gamma}} (1+\gamma) \\
S \propto n^{(1+\gamma)^{-1}}.
\end{split}
\end{equation}

For $\gamma>0$ and $\mu=(1+\gamma)^{-1}$, we find that the number of visited locations ($S$) scales as $\mu<1$. That is, given a certain value of $\gamma$, the model predicts a sublinear scaling, consistent with the empirical observations in the virtual world.

\subsubsection{Visitation Frequency}

A trajectory consists of $n$ jumps, for which at each jump an individual decides to visit a specific location, providing locations $L_1, L_2, L_3... L_S$, where $L_i$ is the $i$th visited location. In the global visitation trajectory, $S(n_i)$ is the first jump number $n_i$ when the location was discovered. As an individual can also re-visit previously visited locations, we can quantify the number of visits to each location by: 

\begin{equation}
    f_i = \frac{m_{i}}{\sum_{j} m_j}
\end{equation}

The sooner a location is discovered, the more likely it is to be visited again, driven by the popularity mechanism: an individual is more likely to visit a popular location. That is, the visitation rank of a location is the same as the discovery order, giving $k(L_i) = S(n_i)$. As the number of visited locations decreases with number of jumps, the probability of finding a location with rank $k(L_i)$ follows $k(L_i) \propto n^{(1+\gamma)^{-1}}$ or $n_{i} \propto k(L_i)^{1+\gamma}$. 

We can write the number of visits to $i$ location as, 

\begin{equation}
    \frac{d m_i}{dn} = \pi(m_i) = \frac{m_i}{\sum_j{m_j}} = \frac{m_i}{n}.
\end{equation}

The solution for the above is $m_i = C_i n$. After solving the equation based on initial conditions, $m_i(C_i) = C_i n_i = 1$, we have $C_i = 1/n_i$. That is, 

\begin{equation}
    m_i = \frac{n}{n_i} \equiv n_{i}^{-1} \equiv k(L_i)^{-(1+\gamma)}.
\end{equation}

In the \textit{EPR} model, the visited location depends on the individual visitation frequencies, while in the \textit{m-EPR} model each individual selects to visit a location based on the location's global visitation frequency. That is, for \textit{m-EPR} model, we have $\pi_j(i) = \pi(i) = \frac{n_i}{\sum_k n_k}$, representing the visitation probability of all agents. Substituting this probability in equation (6) we recover, $n_i \propto k(L_i)^{-(1+\gamma)}$. In summary, we find that the \textit{m-EPR} model analytically recovers the empirically observed power law in visitation frequency with $\alpha^{M}>1$. 

\subsection{Numerical simulations}

\begin{figure}[ht]
\centering
\includegraphics[width=\linewidth, height = 20em, keepaspectratio]{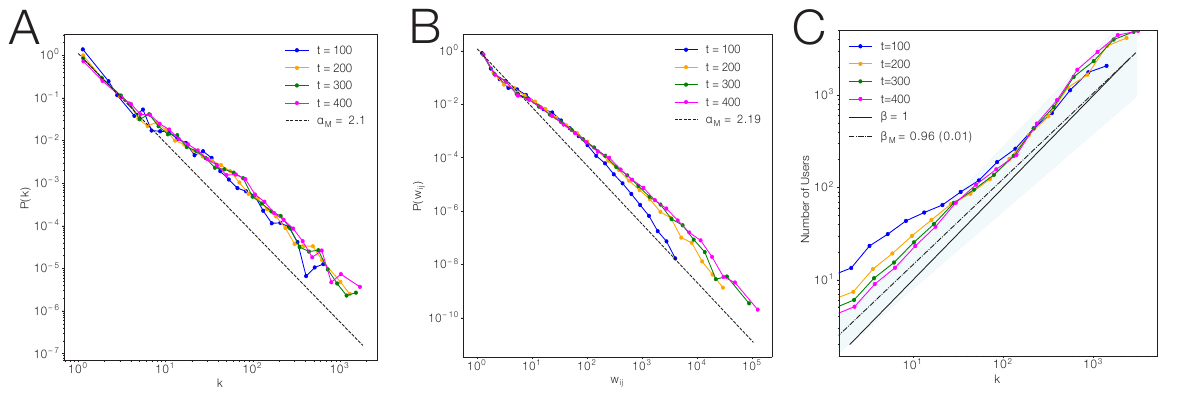}
\caption{\textbf{The effect of time evolution in model predictions.} We show the network characteristics at different time steps: $t=100$, $t=200$, $t=300$, and $t=400$. \textbf{(A)} Degree distribution of the network $P(k) \propto k^{-\alpha}$. \textbf{(B)} Link weight distribution of the network $P(w_{ij}) \propto w_{ij}^{-\alpha}$. \textbf{(C)} Number of visitors and their corresponding degree $N_S \propto S^{\beta}$.}
\label{fig:si_mepr_stat_cond}
\end{figure}

\begin{figure}[ht]
\centering
\includegraphics[width=\linewidth, height = 20em, keepaspectratio]{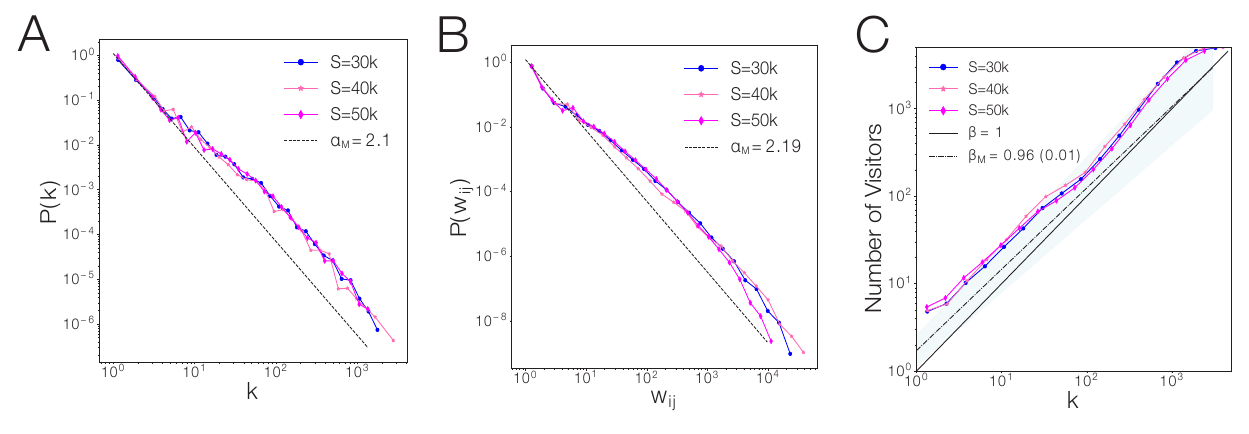}
\caption{\textbf{Impact of location size on model predictions.} We show the model simulation at different metaverse sizes: $S=30,000$, $S=40,000$, $S=50,000$. \textbf{(A)} Degree distribution of the network $P(k) \propto k^{-\alpha}$. \textbf{(B)} Link weight distribution of the network $P(w_{ij}) \propto w_{ij}^{-\alpha}$. \textbf{(C)} Number of visitors and their corresponding degree $N_S \propto S^{\beta}$. The results show that the model results are consistent across different system sizes.}
\label{fig:si_mepr_n_locs_cond}
\end{figure}

We conduct numerical agent-based simulations to predict the individual patterns and the mobility network. The simulations were conducted on a population of $N=5,000$ agents and $S=20,000$ locations. To test the effect of different model specifications on the predictions. We conduct the simulations for different time lengths ($t=100$, $t=200$, $t=300$, $t=400$), and find that the \textit{m-EPR} model is able to capture temporal mobility patterns, revealing stability in the predictive model (Supplementary Fig. \ref{fig:si_mepr_stat_cond}). We also simulate different system sizes $S=20,000$, $S=30,000$, and $S=40,000$ and find that the model predictions are consistent with location size (Supplementary Fig. \ref{fig:si_mepr_n_locs_cond}). 

\begin{figure}[ht]
\centering
\includegraphics[width=\linewidth, height = 20em, keepaspectratio]{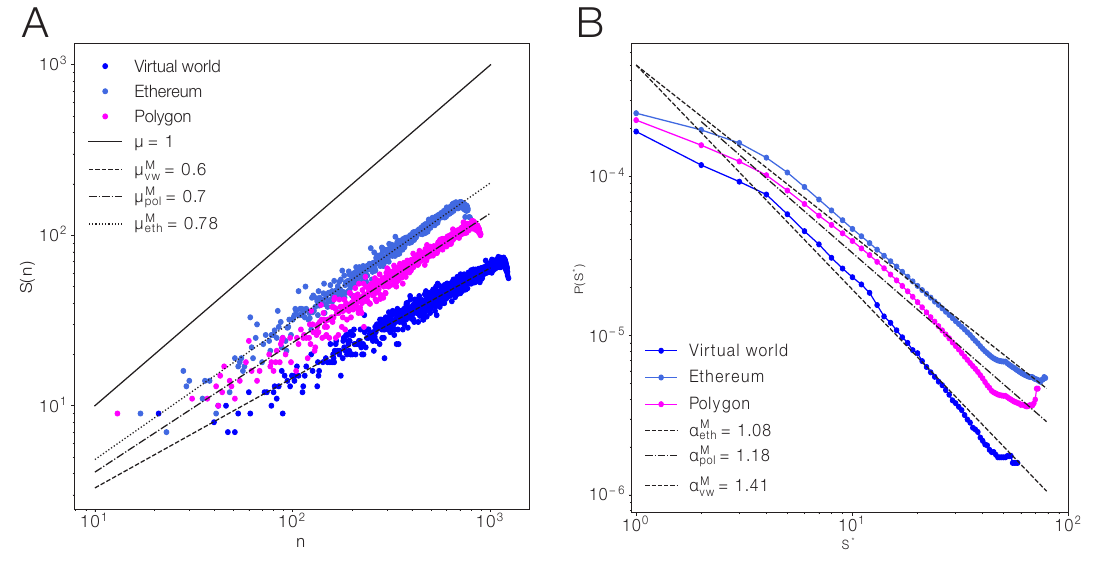}
\caption{\textbf{Individual mobility patterns.} We consider the model parameters from empirical data and simulate individual trajectories. We estimate the observed individual mobility patterns. \textbf{(A)} The number of visited locations $S(n)$ as a function of number of movements $n$. We find that $S(n) \propto n^{\mu^{M}}$, where $\mu^{M}_{vw} = 0.6$, $\mu^{M}_{pol} = 0.7$, and $\mu^{M}_{eth} = 0.78$. As $0 < \mu^{M} <1$, the model uncovers the sub linear exploration patterns observed in the empirical data (key observation (1)). \textbf{(B)} The visitation frequency at different locations. We rank the list of most visited locations, $S^*$ and calculate the number of visitations made to that specific location. We find that $P(S^*) \propto S^{*-\alpha}$, where $\alpha^{M}_{vw} = 1.41$, $\alpha^{M}_{eth} = 1.08$, and $\alpha^{M}_{pol} = 1.18$. As $\alpha^{M}>1$, the model uncovers the power law patterns in visitation frequency (key observation (2)).}
\label{fig:si_mepr_ind_mob}
\end{figure}

Finally, we find that the \textit{m-EPR} model accurately recovers the individual mobility patterns (Supplementary Fig. \ref{fig:si_mepr_ind_mob}): the likelihood the visit new locations decreases, $S(n) \propto n^{\mu^M}$ with $\mu^M < 1$ and $P(S^{*}) \propto S^{*\alpha^M}$ with $\alpha^M>1$, i.e. key observations (1) and (2). We also find that the mobility model recovers the fat-tailed degree distribution and the heterogeneous flows between locations in the network, consistent with the empirical observations (Supplementary Table \ref{tab:model_exponents_si}). 

\clearpage

\begin{table}[ht]
\centering
\caption{\textbf{Modeling metaverse mobility.} We show the exponents from empirical data for Virtual World, Ethereum, and Polygon system, and compare it to the model predictions by \textit{EPR} and \textit{m-EPR} models.}
\begin{tabular}{ l*{5}{c} }
\toprule
\hspace*{2cm} &
\multicolumn{3}{c}{Empirical Observations} &
\multicolumn{2}{c}{Model Predictions}\\
\midrule
\multicolumn{2}{c}{} &
\multicolumn{2}{c}{Network Mobility}\\
\cmidrule(lr){3-4}
& \multicolumn{1}{c}{Virtual World} &
\multicolumn{1}{c}{Ethereum} &
\multicolumn{1}{c}{Polygon} &
\multicolumn{1}{c}{EPR} &
\multicolumn{1}{c}{m-EPR} \\
\midrule
$P(S^*) \propto S^{-\alpha}$ & 1.35 & 1.05 & 1.39 & 1.42 $\pm$ 0.03 & 1.41 $\pm$ 0.02\\
$S(n) \propto n^{\mu}$ & 0.52 & 0.61 & 0.52 & 0.7 $\pm$ 0.01& 0.6 $\pm$ 0.004 \\
\midrule
$P(k) \propto k^{-\alpha}$ & 1.98 & 2.9 & 2.4 & -  & 2.1 $\pm$ 0.06\\
$P(w_{ij}) \propto w_{ij}^{-\alpha}$ & 2.18 & 2.85 & 2.5 & - & 2.19 $\pm$ 0.03\\
$N(S) \propto k^{\beta}$ & 1.05 & 1.05 & 1.13 & - & 0.96 $\pm$ 0.03\\
\bottomrule
\end{tabular}
\label{tab:model_exponents_si}
\end{table}

\end{document}